\documentclass[pra,reprint,doublecolumn,superscriptaddress, nobalancelastpage,dvipsnames]{revtex4-2}
\usepackage[a4paper]{geometry}
\newgeometry{top=1.25cm,right=1cm,left=1.25cm,bottom=1.25cm}
\usepackage[dvipsnames]{xcolor}
\usepackage{amsmath,amssymb,amsfonts}
\usepackage{graphicx}
\usepackage{float}
\usepackage[caption=false,font=normalsize,labelfont=sf,textfont=sf]{subfig}
\usepackage{textcomp}
\usepackage{braket}
\usepackage{diagbox}
\usepackage{array}
\usepackage{booktabs}
\usepackage{tabularx}
\newcolumntype{M}[1]{>{\centering\arraybackslash}m{#1}}
\usepackage{tikz}
\usetikzlibrary{quantikz}
\usepackage[linesnumbered,ruled,vlined]{algorithm2e}

\SetCommentSty{mycommfont}
\SetKwInput{KwInput}{Input}                
\SetKwInput{KwOutput}{Output}              
\newtheorem{assumption}{Assumption}
\newtheorem{theorem}{Theorem}

\usepackage[a4paper]{geometry}
\newgeometry{top=1.25cm,right=1cm,left=1.5cm,bottom=1.25cm}
\usepackage[colorlinks=true, urlcolor=blue, pdfborder={0 0 0}]{hyperref}
\hypersetup{%
linkcolor=blue,
citecolor=BrickRed
}

\begin{document}

\title{\Large{\textbf{Hybrid Quantum-Classical Generative Adversarial Network for High Resolution Image Generation}}}

\author{Shu Lok Tsang}
\affiliation{School of Computing and Information Systems, Melbourne School of Engineering, The University of Melbourne, Parkville, 3010, VIC, Australia}

\author{Maxwell T. West} 
\affiliation{ School of Physics, The University of Melbourne, Parkville, 3010, VIC, Australia}

\author{Sarah M. Erfani}
\affiliation{School of Computing and Information Systems, Melbourne School of Engineering, The University of Melbourne, Parkville, 3010, VIC, Australia}

\author{Muhammad Usman} 
\email{musman@unimelb.edu.au}  
\affiliation{ School of Physics, The University of Melbourne, Parkville, 3010, VIC, Australia}
\affiliation{ Data61, CSIRO, Clayton, 3168, VIC, Australia}

\maketitle

\onecolumngrid

\noindent
\textcolor{black}{
\normalsize{\textbf{Quantum machine learning (QML) has received increasing attention due to its potential to outperform classical machine learning methods in problems pertaining classification and identification tasks. A subclass of QML methods is quantum generative adversarial networks (QGANs) which have been studied as a quantum counterpart of classical GANs widely used in image manipulation and generation tasks. The existing work on QGANs is still limited to small-scale proof-of-concept examples based on images with significant down-scaling. Here we integrate classical and quantum techniques to propose a new hybrid quantum-classical GAN framework. We demonstrate its superior learning capabilities by generating $28 \times 28$ pixels grey-scale images without dimensionality reduction or classical pre/post-processing on multiple classes of the standard MNIST and Fashion MNIST datasets, which achieves comparable results to classical frameworks with three orders of magnitude less trainable generator parameters. To gain further insight into the working of our hybrid approach, we systematically explore the impact of its parameter space by varying the number of qubits, the size of image patches, the number of layers in the generator, the shape of the patches and the choice of prior distribution. Our results show that increasing the quantum generator size generally improves the learning capability of the network. The developed framework provides a foundation for future design of QGANs with optimal parameter set tailored for complex image generation tasks.}}}

\vspace{30pt}

\twocolumngrid

\section{Introduction}
\label{sec:introduction}

\normalsize

Generative adversarial networks (GANs) are one of the best examples of deep learning success in generative learning \cite{gan}. It consists of a generator and a discriminator competing against each other, where the generator attempts to generate realistic data (such as images) while the discriminator attempts to differentiate between real and generated data. Ultimately, the goal of the framework is to have the generator distribution replicate the training data distribution, which is mathematically equivalent to minimising the Jensen-Shannon (JS) divergence between them \cite{gan}. GANs have been deployed in many application areas such as image generation \cite{dcgan}, future prediction in videos \cite{gan-future-pred}, text to image synthesis \cite{gan-tti}, image-to-image translation \cite{gan-iti}.  Despite their empirical success, GANs suffer from a variety of problems during training in practice, namely vanishing gradients, mode collapse and a lack of stopping criteria \cite{gan-principled, wgan}. There have been many proposed improvements to tackle these problems. One particular proposal is the Wasserstein GAN (WGAN) \cite{wgan}, where the training is reformulated to minimise the Wasserstein distance instead. The WGAN framework has demonstrated empirically that it can effectively tackle the aforementioned problems. However as with other classical GANs, training on complex datasets require large networks and amounts of computational resources.

\begin{figure*}[ht]
    \begin{center}
        \includegraphics[width=\textwidth]{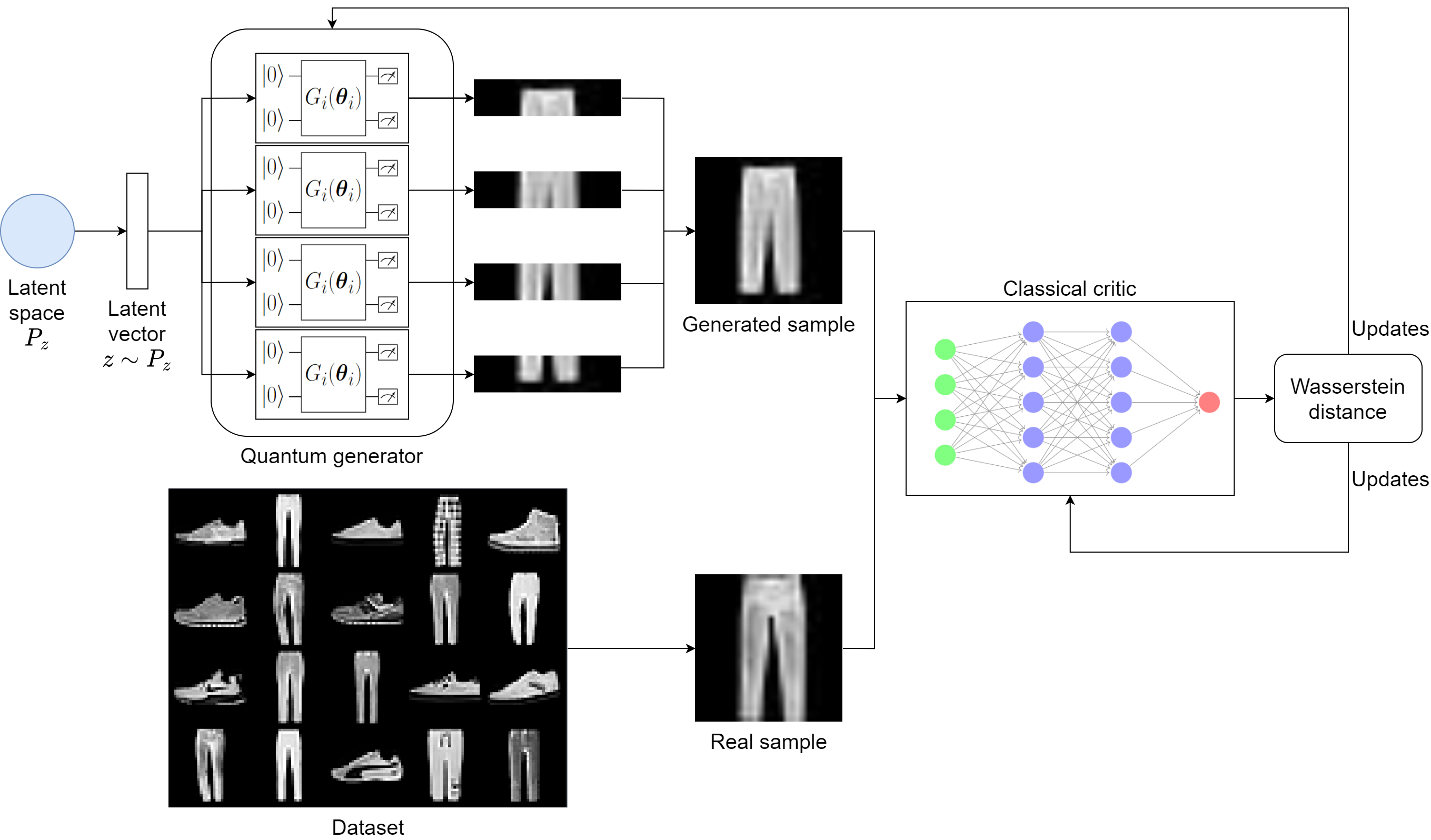}
        \caption{\label{fig:1} \textbf{Overview of PQWGAN framework.} Our framework is identical to the WGAN-GP framework, with the difference being the fake images are now being generated by a quantum generator. The framework operates as follows. First, a latent vector $\boldsymbol{z}$ is sampled from the latent space and is encoded using $R_Y$ rotations as $\ket{\boldsymbol{z}}$ in each sub-generator of the quantum generator. The relevant qubits are measured at the end of the circuit and post-processed to create a patch of the image. The patches are stitched together to form a complete generated image. The generated and real images are then passed to the critic which estimates the Wasserstein distance between the generator and real distribution. Finally, these statistics are used to update the generator and the critic.}
    \end{center}
\end{figure*}

The emergence of quantum computing as a new computing paradigm has led to quantum algorithms that show great promise to solve many of the computationally hard problems in computer science, such as Shor’s algorithm for efficient prime number factorisation \cite{shor}. Since quantum mechanics can generate counter intuitive patterns in data, it is believed that quantum computers can recognise classically challenging patterns \cite{qml}. These promises have led to the development of quantum machine learning (QML), where quantum algorithms are used to improve existing machine learning techniques. Benefits that QML brings include providing speed-ups in training time \cite{qsvm, q-clustering, q-nearest} or obtaining better model performance \cite{qbm, west2022benchmarking, vqc, pqc-expressive}. This may allow QML models to complement or even replace classical methods in the future as the complexity of the tasks continuously increases. However, we are currently in the noisy intermediate-scale quantum (NISQ) era of quantum computation \cite{nisq}. Reliably executing large scale quantum algorithms on current quantum hardware is difficult due to engineering challenges such as noise mitigation. Under these restrictions, much QML research has been focused on quantum algorithms that are compatible with NISQ devices \cite{qml}, such as developing hybrid quantum classical solutions with parameterised quantum circuits (PQCs) \cite{pqc}.

The intersection of quantum computing and GANs have led to the birth of a new research direction known as quantum generative adversarial networks (QGANs) \cite{qgan, qgan-sim}, which aims to push forward generative learning. Currently, QGANs are still in its infancy, and many proposed frameworks deal with low-dimensional data such as simple probability distributions \cite{qgan-dist-1, qgan-dist-2, qgan-drug}. In the realm of image generation, QGANs built using quantum generators have only been able to generate low resolution images in \cite{p-qgan} or require dimensionality reduction with principal component analysis (PCA) \cite{QuGAN, iqgan}. Also, there exists an important knowledge gap in QGANs on how varying different parameters within the quantum generator affects the performance and output quality of QGANs. In previous works, the evaluation of the QGANs are conducted on low-dimensional data, where the images have either been compressed to a lower dimensional space in \cite{QuGAN} and \cite{iqgan}, or are from a synthetic $2 \times 2$ pixels dataset in \cite{p-qgan}. This restricts their application for realistic problems and the scope of the acquired understanding is also limited.

In this paper, we aim to bridge the gap between classical and quantum GANs by generating high dimensional data in the form of images. Specifically, we propose a new hybrid quantum-classical framework as shown in Fig. \ref{fig:1}, which we call the patch quantum Wasserstein GAN (PQWGAN). Our framework leverages the theoretical benefits that the WGAN-GP \cite{wgan-gp} brings to improve the patch strategy QGAN \cite{p-qgan}. The patch strategy QGAN splits the output image generated into different patches, each generated by a separate quantum circuit. This is useful in the NISQ era, where splitting up the output can reduce the quantum resources required. On the other hand, the WGAN-GP is an extension to the WGAN that has improved convergence properties owing to the use of a gradient penalty for regularisation. Individually, the patch QGAN with the GAN framework \cite{p-qgan} and the WGAN-GP framework with a single PQC quantum generator (see Section \ref{section:number-of-patches}) are both unable to generate high resolution images. However, combining these two ideas our PQWGAN is capable of generating high resolution images without dimensionality reduction, which was not possible using previous QGANs directly. 

 First, we numerically demonstrate the viability of the framework by applying it to learn to generate full resolution $28 \times 28$ pixels images from the standard MNIST \cite{mnist} and Fashion MNIST (FMNIST) \cite{fmnist} datasets. To the best of our knowledge, this is the first demonstration of a QGAN that uses quantum circuits as a generator that can successfully generate images without dimensionality reduction or classical pre/post-processing at this scale. Second, we gain a deeper understanding of the effects that varying different quantum generator parameters have on output quality by experimenting directly on the $28 \times 28$ pixels images.  Parameters explored include the number of patches, qubits, layers, shape of patches and choice of prior. Simulations provide the crucial insight that increasing the generator size in general correlates with better output quality. Our results demonstrate that our framework has the potential to serve as a foundation for future QGAN research on more complex tasks.

This paper is organised as follows. First, we go over some preliminaries in Section \ref{section:prelim} and related work on QGANs in Section \ref{section:related}. Next, we introduce our novel PQWGAN framework in Section \ref{section:pqwgan} and our experimental setup in Section \ref{section:setup}. Then, we put our PQWGAN to the test by applying it to generate images of MNIST and FMNIST in Section \ref{section:image-generation} and evaluating the effects of different parameters in Section \ref{section:effects}. Finally, we conclude our findings and propose future directions in Section \ref{section:conclusion}.

\section{Preliminaries}
\label{section:prelim}

\subsection{Generative adversarial networks}

GANs were first proposed in \cite{gan}. The framework consists of a discriminator and a generator that compete in an adversarial game. The generator and discriminator can in theory be any machine learning model, but are commonly neural networks due to its empirical success. The generator $G$ takes an input noise vector $\boldsymbol{z}$ sampled from some distribution $P_{\boldsymbol{z}}$ (e.g., Gaussian) and produces an output. The goal is to have the learned distribution $P_G$ match the real distribution $P_{data}$. On the other hand, the discriminator $D$ takes an input $\boldsymbol{x}$ and outputs the probability it believes $\boldsymbol{x}$ originated from $P_{\boldsymbol{z}}$. If the probability is greater than one half, the input is classified as originating from the real data and vice versa. The goal of $D$ is to maximise the probability of assigning the correct labels, while the goal of $G$ is to produce samples that confidently fools $D$. The objective of the GAN training can be expressed in terms of a zero-sum game,
\begin{equation}
    \label{eq:GAN}
    \min_G \max_D \mathbb{E}_{\boldsymbol{x} \sim P_{data}} [\log D(\boldsymbol{x})] 
    + \mathbb{E}_{\boldsymbol{z} \sim P_{\boldsymbol{z}}} [\log (1 - D(G(\boldsymbol{z})))].
\end{equation}

\subsection{Wasserstein Generative adversarial networks}

As briefly mentioned in Section \ref{sec:introduction}, GANs suffer from a variety of problems during training due to the mathematical properties of minimizing the JS divergence. To mitigate these issues, there have been efforts to reformulate GAN training with completely different objectives in order to obtain better theoretical guarantees. One of the most successful framework is the Wasserstein GAN (WGAN) \cite{wgan}. The WGAN minimises the Wasserstein distance, and the value function of the WGAN is,
\begin{equation}
    \label{eq:WGAN}
    \min_G \max_{D \in \mathcal{D}} \mathbb{E}_{\boldsymbol{x} \sim P_{data}} [D(\boldsymbol{x})] 
    - \mathbb{E}_{\boldsymbol{z} \sim P_{\boldsymbol{z}}} [D(G(\boldsymbol{z}))],
\end{equation}
where $\mathcal{D}$ is the set of 1-Lipschitz functions. Instead of having a discriminator that produces a binary output as in GANs, the discriminator now outputs a score which is interpreted as the Wasserstein distance between $P_G$ and $P_{data}$. Hence, the discriminator is known as a critic instead. Minimising the Wasserstein distance exhibits much nicer theoretical guarantees than minimising the JS divergence, and is shown to converge in many instances where the JS divergence fails to do so \cite{wgan}. First, the gradient of the critic with respect to the input is much better behaved than that of the discriminator in GANs, allowing the generator to be trained more easily and the critic to be trained to optimality without having to deal with vanishing gradients. Next, the WGAN has shown empirical evidence of being able to avoid mode collapse, as the authors were able to train various discriminator and generator architectures that were previously hard to train successfully. Also, since the WGAN value function provides an estimate of the Wasserstein distance, it has empirically observed to be correlated with sample quality of the generator. Hence, this can be used as a stopping condition in WGAN training.

To enforce the 1-Lipschitz constraint, \cite{wgan} proposed to clip the gradients of the each critic parameter within a fixed range such as $[-0.01, 0.01]$. However, the choice of the clipping range poses new problems. If the gradient magnitude is large, it can take a long time for the critic to reach optimality, but if the magnitude is small it can also easily lead to vanishing gradients. Instead, \cite{wgan-gp} proposed the WGAN-GP, where a gradient penalty is used instead to enforce the 1-Lipschitz constraint. The new value function is,
\begin{multline}
    \label{eq:WGAN-GP}
    \min_G \max_{D \in \mathcal{D}} \mathbb{E}_{\boldsymbol{x} \sim P_{data}} [D(\boldsymbol{x})]
    - \mathbb{E}_{\boldsymbol{z} \sim P_{\boldsymbol{z}}} [D(G(\boldsymbol{z}))]
    - \\ \lambda \mathbb{E}_{\boldsymbol{\hat{x}} \sim P_{\boldsymbol{\hat{x}}}} [(||\nabla_{\boldsymbol{\hat{x}}} D(\boldsymbol{\hat x})||_2 - 1)^2],
\end{multline}
where $\lambda$ is a constant and $P_{\hat{x}}$ is a distribution sampled uniformly in between $P_{data}$ and $P_{G}$. In the WGAN framework, the optimal critic has unit gradient norm for straight lines between $P_{data}$ and $P_G$. Hence, by enforcing this condition in Eq. (\ref{eq:WGAN-GP}), the critic is able to be trained to optimality without vanishing gradients. This is supported by the fact that the WGAN-GP framework was applied to successfully train many random variations of the DCGAN architecture, such as having different activation functions, depth, use of batch normalisation and filter count \cite{wgan-gp}. Compared to the GAN framework, the WGAN-GP framework is able to successfully train a significantly larger portion of these random architectures to some minimum Inception score \cite{improved-gan} (which quantitatively measures the output variety and quality of a GAN) on the $32 \times 32$ pixels ImageNet dataset.

\section{Related work on quantum generative adversarial networks}
\label{section:related}

The notion of a QGAN was first introduced theoretically in \cite{qgan}, and demonstrated to be viable numerically in \cite{qgan-sim}. In general, QGANs can take in either quantum or classical data. For classical data, \cite{qgan} claims that although there are no guarantees for quantum advantage, it is reasonable to expect that the quantum GAN can learn the data distribution in less time due to efficient quantum algorithms to solve linear equations such as the Harrow-Hassidim-Lloyd (HHL) algorithm \cite{hhl}. However, early methods focus on generating relatively simple low-dimensional distributions. As such, many of the proposed QGAN frameworks provide limited use for high resolution image generation in the NISQ era.

\subsection{Quantum generative adversarial networks for image generation}

Given a quantum computer with $n$ qubits and the task of generating a $M$ dimensional output, \cite{p-qgan} suggested the batch and patch strategy for QGANs in the NISQ era. The patch strategy is useful for the case where $n < \lceil \log M \rceil$, which is likely the case in higher resolution image generation on NISQ devices. In the patch QGAN, the generator is composed of $k$ quantum circuits that are each sampled to generate a patch of the image, while the discriminator can be either a classical or quantum classifier. The resulting patches are then stitched together to form the final image. The patch QGAN was able to generate 8 × 8 pixels images of handwritten digits of 0s and 1s by training on the optical recognition of handwritten digits dataset \cite{digits} on both simulations and a superconducting quantum computer. Although this approach successfully generated images of handwritten digits, the quality of the images were quite low.

Another method explored to generate images is the state fidelity based QGAN (QuGAN) \cite{QuGAN}. The core of the framework is a swap test to measure the fidelity between the discriminator and generator state for the loss function. Recently, the IQGAN \cite{iqgan} was proposed as an extension to the QuGAN framework. It features a new trainable classical to quantum encoder to embed classical data and a more compact quantum generator that avoids costly two qubit gates. Both frameworks carried out experiments using simulations and real devices on a subset of the MNIST dataset compressed using PCA, and were both able to successfully generate the target images. However, due to the use of an inverse PCA to generate the images from a low dimensional representation, the images are often quite blurry. Also, since the diversity of the output only stems from the randomness when doing a finite measurement on the generator state, it may be difficult to scale to more complex tasks such as having more digits.

\subsection{Quantum Wasserstein generative adversarial networks}

The idea of a fully quantum version of the Wasserstein GAN (qWGAN) was proposed in \cite{qwgan} and \cite{qwgan-2}.  Both of these works were concerned with the task of learning to generate pure and mixed states using the Wasserstein distance with PQCs. In both cases, simulations showed that the frameworks are able to learn the target states and converge to a high fidelity quickly. However, since their framework is designed to work on quantum data, it cannot be directly applied to image generation, which is what we are interested in.

Another extension of the WGAN is the QWGAN-GP \cite{qwgan-gp}, which considered a hybrid quantum-classical version of the WGAN-GP. In this framework, the generator is a single PQC that takes a latent vector as input, while the remaining components of the QWGAN-GP are identical to the WGAN-GP. Experiments on the credit card fraud dataset \cite{credit-card} showed that the QWGAN-GP has comparable performance to a fully connected WGAN-GP architecture on anomaly detection while having less trainable parameters. However, the results indicate that the dataset is too simple, as both the classical and quantum networks converges to the optimum with a low dimensional latent vector and low depth for the generators.

\section{PQWGAN framework}
\label{section:pqwgan}

In this section, we present the patch quantum Wasserstein GAN (PQWGAN) framework for generating high resolution images on NISQ devices. A comparison of our work to existing QGANs for image generation is shown in Appendix \ref{section:details}. The PQWGAN integrates the patch method for image generation on NISQ devices \cite{p-qgan} and the WGAN-GP \cite{wgan-gp}. We choose to use WGAN-GP due to its improved convergence properties compared to weight clipping. Furthermore during initial explorations with QGANs, we noticed that when applying the GAN loss function as in Eq. (\ref{eq:GAN}), the QGANs exhibited unstable behaviour, and were hard to train. Hence, this challenge further motivated us to use a theoretically stable method like WGAN-GP. In our case, the setup is the same as in WGAN-GP, but instead of a classical generator, we replace it with a patch quantum generator as in \cite{p-qgan}. The overall architecture of the PQWGAN is shown in Fig. \ref{fig:1}.

\subsection{Structure of quantum generator}
\label{section:pqwgan-generator}

\begin{algorithm}[ht]
    \DontPrintSemicolon

    \KwInput{Image dimensions $H \times W$, number of ancilla qubits $A$, number
    of data qubits $D$, number of sub-generator layers $L$, number of patches
    $P$, generator parameters $\boldsymbol{\theta} = [\theta_1, ..., \theta_P]$,
    latent variable $\boldsymbol{z}$.}

    \For{$i=1,...,P$}{
        $\ket{\psi_i} \leftarrow \mathcal{U}_{i,L,\theta_i} \ket{\boldsymbol{z}}$ 

        $\rho_D \leftarrow \text{Tr}_A \left(\frac{(\ket{0} \bra{0})^{\otimes
        A} \otimes \mathbb{I} \ket{\psi_i} \bra{\psi_i}}
        {\braket{\psi_i | (\ket{0} \bra{0})^{\otimes A} \otimes
        \mathbb{I} | \psi_i}} \right)$

        measure $\rho_D$ in computational basis to obtain $G_i(\boldsymbol{z})
        \leftarrow [p(0), ..., p(2^D - 1)]$

        $G'_i(\boldsymbol{z}) \leftarrow \frac{G_i(\boldsymbol{z})}{\max (G_i(\boldsymbol{z}))}$

        discard excess pixel values to obtain $G''_i(\boldsymbol{z}) \leftarrow
        G'_i(\boldsymbol{z})[:\frac{HW}{P}]$
    }

    $G(\boldsymbol{z}) \leftarrow [G''_1(\boldsymbol{z}), ..., G''_P(\boldsymbol{z})]^T$

    \Return{$G(\boldsymbol{z})$}

\caption{Algorithm to generate an image from the patch quantum generator.}
\label{algorithm:quantum-gen}
\end{algorithm}

\begin{figure*}[ht]
    \centering
    $
    \begin{quantikz}
        \lstick[wires=5]{$\ket{0}^{\otimes N}$} & \gate{R_Y(z_1)} & \gate{R(\phi_{i,1}, \theta_{i,1}, \omega_{i,1})} 
        \gategroup[wires=5,steps=4,style={inner sep=6pt}]{Repeated $L$ times} &
        \ctrl{1} & \qw & \qw & \qw & \meter{} \rstick[wires=2]{$A$ ancilla
        qubits} \\
        & \gate{R_Y(z_2)} & \gate{R(\phi_{i,2}, \theta_{i,2}, \omega_{i,2})} &
        \targ{} &\ctrl{1}& \qw & \qw & \meter{} \\
        \wave&&&&&&& \rstick[wires=3]{$D$ data qubits}\\
        & \gate{R_Y(z_{N-1})} & \gate{R(\phi_{i,N-1}, \theta_{i,N-1},
        \omega_{i,N-1})} & \qw & \targ{} &\ctrl{1}& \qw & \meter{} \\
        & \gate{R_Y(z_N)} & \gate{R(\phi_{i,N}, \theta_{i,N}, \omega_{i,N})} \qw
        & \qw & \qw & \targ{} & \qw & \meter{}
    \end{quantikz}
    $
    \caption[Quantum circuit of a sub-generator]{\textbf{Quantum circuit of a sub-generator}. First, each component of the latent vector is encoded into the rotation angle of a $R_Y$ gate. Then, the state passes through $L$ layers of arbitrary parametrised rotations $R$ and CNOTs using the hardware efficient structure. The subscripts $i,j$ of the parameters of the $R$ gates refer to the $i$-th layer and the $j$-th qubit. The ancilla qubits are measured to perform a non-linear transformation on the state. In this paper, we use only one ancilla qubit and pick the resulting state where the ancilla is 0. Finally, the data qubits are measured to form a patch.}
    \label{figure:sub-generator}
\end{figure*}
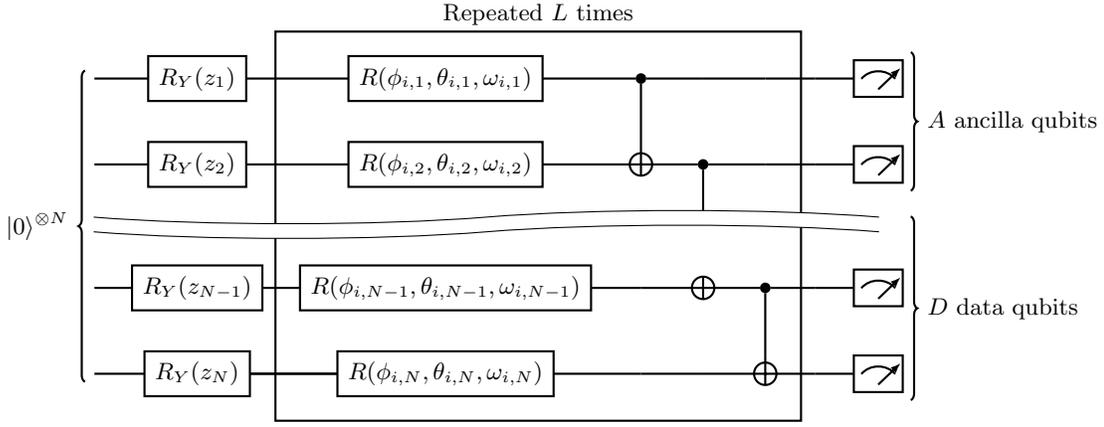

Algorithm \ref{algorithm:quantum-gen} shows the process of generating an image using the quantum generator is shown in. The generator is composed of $P$ quantum circuits, which we will refer to as sub-generators. The circuit for a sub-generator is shown in Fig. \ref{figure:sub-generator}. Each sub-generator $\mathcal{U}_{i, L, \theta_i}$ is a PQC with $N$ qubits and $L$ repeated layers of parameterised gates. The sub-generator is split into two components, the $A$ ancilla qubits, and the $D$ data qubits. The latent vector is first encoded directly into the rotation angles of $R_Y$ gates. Then, the state is transformed through the layers of the PQC. At the end of the quantum circuit, the ancilla qubits are measured to perform a non-linear transformation of the latent vector. Then, we measure the required number of data qubits in the computational basis and normalise their amplitudes to obtain valid pixel values. Finally, we stack these patches together to obtain a valid image. A more detailed discussion of the implementation of the quantum generator can be found in Appendix \ref{section:details}.

\subsection{Structure of critic}

The critic in this case is the same as in WGAN-GP, which is a classical neural network. We believe that a classical neural network would be better for the PQWGAN framework as it is developed for NISQ devices in mind. Recall that the critic is responsible for taking in an image and outputting a real value that serves as an estimate for the Wasserstein distance. This poses several problems for a quantum critic. First, we would have to load high dimensional data into a quantum circuit, which is hard to do in practice due to the amount of quantum resources required. Secondly, the learning process of quantum circuits are not as well understood as classical neural networks. There are limited solutions to problems that are frequently encountered in quantum learning such as barren plateaus \cite{barren-plateaus}.

\subsection{Training objective}

We adopt the objective of WGAN-GP defined in Eq. (\ref{eq:WGAN-GP}) to train the PQWGAN. A subtle difference to WGAN-GP is that the output from the generator originates from multiple sub-generators as detailed in Section \ref{section:pqwgan-generator}. The use of this objective is motivated by the empirical observation that WGAN-GP can be used to train a variety of different architectures successfully with minimal hyperparameter tuning. We view this as an important benefit for the NISQ era since quantum resources are scarce. Although quantum models can be prototyped using quantum simulators, the difficulty of simulating quantum circuits means that searching for optimal hyperparameters is a time consuming and tedious task. Until we can more efficiently execute quantum circuits, fine grained hyperparameter tuning will most likely be out of reach for larger models. Hence, this objective will in principle allows us to train QGANs in general with a greater success rate.

\subsection{Training algorithm}

\begin{algorithm}[ht]
    \DontPrintSemicolon
      
    \KwInput{Gradient penalty coefficient $\lambda$, critic iterations per
    generator iteration $n_C$, number of epochs $n_{epochs}$, batch size $m$,
    Adam hyperparameters $\eta_1, \eta_2, \beta_1, \beta_2$.}

    initialise critic parameters $\boldsymbol{w}$, sub-generator parameters $\boldsymbol{\theta}$

    \For{$epoch=1,...,n_{epochs}$}{
        \For{$t=1,...,n_C$}{
            \For{$i=1,...,m$}{

            Sample real data $\boldsymbol{x} \sim \mathbb{P}_{data}$, latent
            variable $\boldsymbol{z} \sim p_{\boldsymbol{z}}$,
            random number $\epsilon \sim U[0,1]$
        
            $\boldsymbol{x}' \leftarrow  quantum\_generator(\boldsymbol{\theta},
            \boldsymbol{z})$

            $\hat{\boldsymbol{x}} \leftarrow \epsilon \boldsymbol{x} + (1 -
            \epsilon) \boldsymbol{x}'$

            $L_D^{(i)} \leftarrow D(\boldsymbol{x}') - D(\boldsymbol{x}) +
            \lambda (||\nabla_{\hat{\boldsymbol{x}}} D(\hat{\boldsymbol{x}})||_2 - 1)^2$

            }
            
            $\boldsymbol{w} \leftarrow$ Adam $(\nabla_{\boldsymbol{w}}
            \frac{1}{m} \sum_{i=1}^m L_D^{(i)}, \boldsymbol{w}, \eta_1, \beta_1,
            \beta_2)$

        }

        \For{$i=1,...,m$}{
            Sample latent variable $\boldsymbol{z} \sim p_{\boldsymbol{z}}$

            $\boldsymbol{x}' \leftarrow quantum\_generator(\boldsymbol{\theta},
            \boldsymbol{z})$

            $L_G^{(i)} \leftarrow - D(\boldsymbol{x}')$
        }

        $\boldsymbol{\theta} \leftarrow$ Adam $(\nabla_{\boldsymbol{\theta}}
        \frac{1}{m} \sum_{i=1}^m L_G^{(i)}, \boldsymbol{\theta}, \eta_2,
        \beta_1, \beta_2)$      
        
    }
\caption{PQWGAN training algorithm.}
\label{algorithm:PQWGAN}
\end{algorithm}

The training algorithm for PQWGAN follows the WGAN-GP training algorithm, except for the use of quantum generators (see Algorithm \ref{algorithm:PQWGAN}). With the generator now being split into sub-generators, we have to update the parameters of different sub-generators given a loss that is calculated on the entire images. Furthermore, the loss function $\mathcal{L}(\boldsymbol{w}, \boldsymbol{\theta})$ depends on both the critic and generator parameters respectively. Since the training is done in alternating steps, and either the critic or generator is assumed to be fixed while training, we can still use the parameter shift rule \cite{parameter-shift} to compute the gradient. Assuming we have $N_G$ sub-generators with $n$ parameters each, the generator's parameters can be expressed as a vector $\boldsymbol{\theta} = [\boldsymbol{\theta}_1, ..., \boldsymbol{\theta}_{N_G}] = [\theta_{1,1}, ..., \theta_{1,n}, ..., \theta_{N_G,1}, ..., \theta_{N_G, n}]$. Hence, the gradient of the $j$-th parameter of the $i$-th sub-generator with respect to the loss is
\begin{multline*}
    \frac{\partial \langle \mathcal{L}(\boldsymbol{w}, \boldsymbol{\theta}) \rangle}{\partial \theta_{i,j}}
    = \frac{1}{2}(\langle \mathcal{L}(\boldsymbol{w}, [\theta_{1,1}, ..., \theta_{i,j} + \pi/2,..., \theta_{N_G, n}]) \rangle - \\ \langle \mathcal{L}(\boldsymbol{w}, [\theta_{1,1}, ..., \theta_{i,j} - \pi/2,..., \theta_{N_G, n}]) \rangle)
\end{multline*}

\section{Experimental setup}
\label{section:setup}

\subsection{Dataset and libraries}

We pick the publicly available MNIST \cite{mnist} and FMNIST \cite{fmnist} datasets to conduct our experiments on. Although MNIST and FMNIST are simple datasets for classical GANs, they are still a considerable step up in terms of complexity to previous studies of QGANs especially with the full $28 \times 28$ resolution. Due to resource limitations, we will only be using the first 1000 samples of every class that we include in our training set to ensure that the training process can be carried out in a reasonable amount of time.

All models are implemented in Python3 using PyTorch \cite{pytorch} and PennyLane \cite{pennylane}. PyTorch is a high performance machine learning library while PennyLane is a QML library which provides interfaces to PyTorch. The training of all PQWGANs are simulated without noise using high performance computing resources from the National Computational Infrastructure, Pawsey and the University of Melbourne.

\subsection{Classical critic and generator structures} 

To systematically investigate the performance of the quantum generator, we fix the critic in the PQWGAN to be the same in all of our experiments. The critic is a fully connected network with two hidden layers of 512 and 256 neurons, respectively. Both of these hidden layers have a leaky ReLU activation with a slope of 0.2. The final hidden layer is connected to an output layer of one neuron with no activation to obtain a real valued output. 

To contrast our framework with classical GANs, we compare our results to a WGAN-GP. For consistency, the critic used is the same as in the PQWGAN. The generator is now also a fully connected network with three hidden layers of 256, 512 and 1024 neurons respectively. Again, the hidden layers all have leaky ReLU activations with a slope of 0.2. Finally, the hidden layers are connected to an output layer consisting of the same number of neurons as the output pixels with a $\tanh$ activation, which are then rearranged to form an image. 

Since our experiments are conducted on relatively simple datasets, we opted for a simple architecture across all classical components of our experiments. To validate the capability of the classical parts in learning, we applied the WGAN-GP to learn the full MNIST and FMNIST dataset. In both datasets, the Wasserstein distance converges towards 0, while manually inspecting the outputs confirmed that the generator is indeed learning successfully. This shows that our classical components should not affect the learning capabilities of the PQWGAN.

\subsection{Hyperparameters}

Unless otherwise specified, the hyperparameters for all our experiments are chosen as follows. We follow the default values for the learning process in WGAN-GP \cite{wgan-gp}, where we use the values $\lambda = 10, n_C = 5$ and Adam \cite{adam} for optimisation with hyperparameters $\beta_1 = 0, \beta_2 = 0.9$. We decided on having 28 sub-generators generating 28 patches, so that one patch would correspond to one row of pixels in the image. Each sub-generator also has 1 ancilla qubit. Furthermore, after some hyperparameter tuning, we found that the learning rate for the quantum generator needs to be higher than the classical critic to learn, and we set the learning rate to be 0.01 and 0.0002 for the generator and critic respectively. Also, in our initial explorations the quantum generator was observed to learn quicker when using a uniform prior, so we chose a uniform prior over a Gaussian prior. The uniform prior is restricted to be in the range $[0,1)$ instead of $[-\pi,\pi)$. Although using the latter can cover the whole range of possible rotations in the quantum circuit, we found that it led to poorer learning due to the larger space that the generator has to learn from. 

Due to the time required to simulate the quantum circuits, we use a batch size of 25 to ensure that the generator is sufficiently updated during the training process. In all our experiments, we train the generator for 600 iterations, which is equivalent to processing 3000 batches of data in total. Depending on the number of classes used in our experiments, this corresponds to 37.5 or 25 epochs for the two and three class experiments respectively. We pick this number as it is a considerable number of epochs for training under current resource constraints while also being able to be completed in a reasonable amount of time.

\section{Image generation with PQWGAN}
\label{section:image-generation}

\begin{figure*}[ht]
    \begin{center}
        \includegraphics[width=\textwidth]{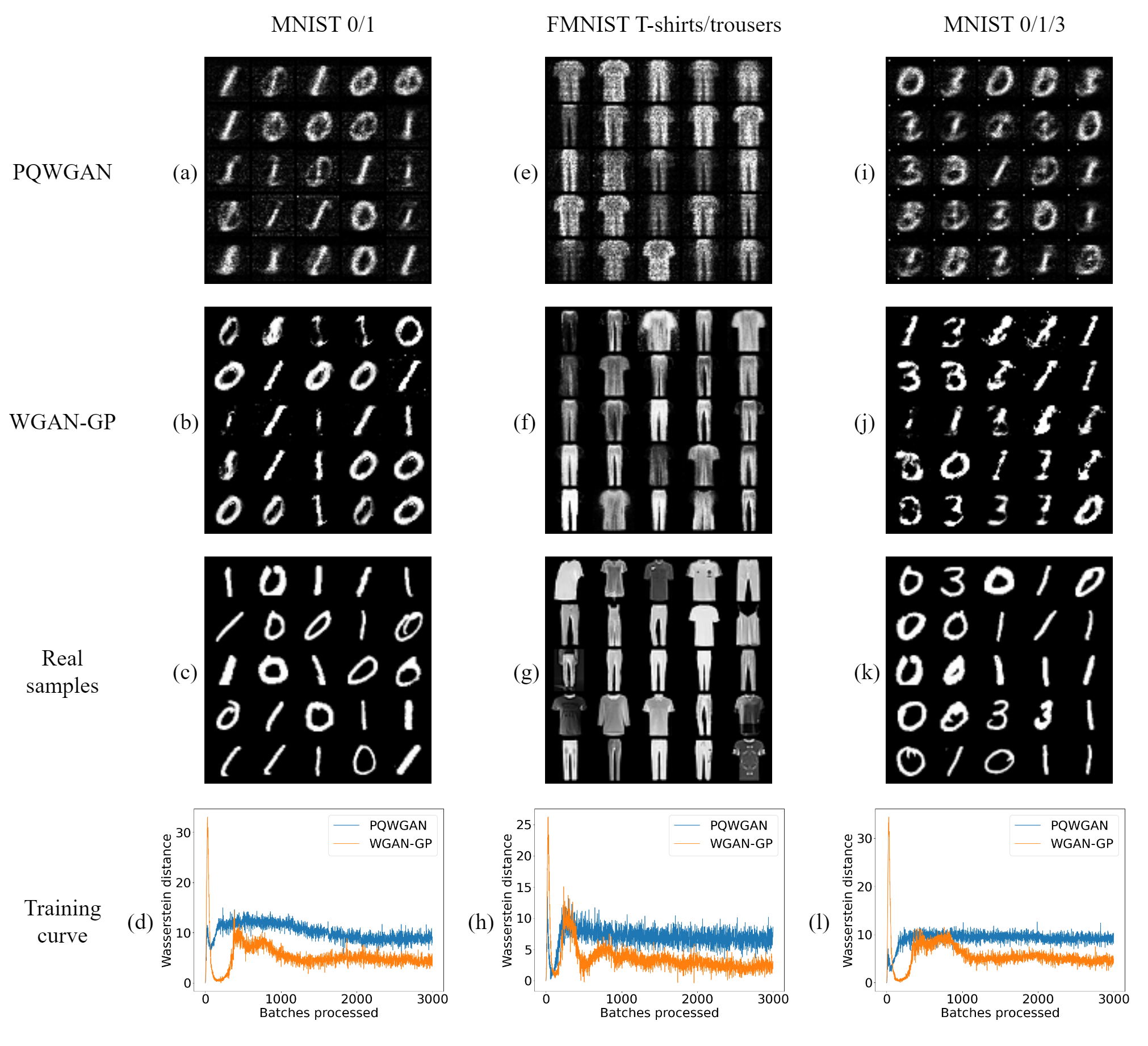}
        \caption{\label{fig:image-gen} \textbf{Random images generated using PQWGAN and WGAN-GP.} The images generated by our PQWGAN are of comparable quality to that using a WGAN-GP albeit having orders of magnitude less trainable parameters. Especially for MNIST 0/1 and FMNIST T-shirts/trousers, the PQWGAN is able to keep up with WGAN-GP in terms of image sharpness. However, for MNIST 0/1/3 both the PQWGAN and WGAN-GP starts to struggle.}
    \end{center}
\end{figure*}

In this section, we apply our PQWGAN framework to generate images of MNIST and FMNIST.  To compare between classical and quantum learning, we repeat each task with a fully classical WGAN-GP with identical hyperparameters except for the use of a Gaussian prior and the learning rate, where it was a constant 0.0002 for both the generator and critic. These exceptions were made to ensure that we are not severely crippling the learning ability of the WGAN-GP. We also sampled from a latent space that has the same number of dimensions as available in the quantum case to ensure that the WGAN-GP cannot exploit extra dimensions in the latent space. Our results are shown in Fig. \ref{fig:image-gen}.

\subsection{Binary MNIST}
\label{section:binary-mnist}

First, we applied the PQWGAN to generate digits 0 and 1 from MNIST. In this task, each sub-generator consists 8 layers of 7 data qubits and 1 ancilla qubit to conserve resources. As such, the latent vector has 8 dimensions.

Looking at Figs. \ref{fig:image-gen}(a) and \ref{fig:image-gen}(b), it is evident that for both the classical and quantum case, the generators are successfully learning to generate images of 0s and 1s. However, in terms of the image quality, there are still imperfections in both cases which allows them to be easily distinguished from the real images. In particular for the quantum case, there are some samples which resemble neither a 0 nor 1 and appears to be a combination of the two, such as row 3 column 3 in Fig. \ref{fig:image-gen}(a). These mixed images can be attributed to the incomplete learning process when the generator has yet not learned a comprehensive mapping for the entire latent space. Due to the limitations of running quantum simulations, 600 generator iterations is a very small amount compared to experiments conducted in the classical case, where the number is around $10^4$ to $10^5$ generator iterations albeit for a more complex task. Furthermore, in classical GAN training, the generator also outputs samples that look like a combination of the different classes in the early stages of training before slowly learning to diversify into the separate digits. The similar behaviour between PQWGAN and WGAN-GP suggests that the problem of having mixed images can be mitigated by training the generator more.

Another imperfection in the quantum case is the fuzziness of the images. Even in images that look plausible, such as row 1 column 3 in Fig. \ref{fig:image-gen}(a), we can still see that the edges are not very sharp. Since the pixel values originate from the amplitudes of the quantum state at the end of the circuit, to generate a completely dark pixel, we require the corresponding amplitude to be 0. However, due to the nature of a highly entangled circuit, it is very difficult for a particular state to have exactly 0 amplitude. This effect is magnified by the post processing step, where the amplitudes are normalised to be in [0, 1]. In terms of image sharpness, classical GANs have an advantage since it is easier for the optimisation process to change the value of a specific pixel.

The fact that the PQWGAN can generate images of comparable quality to WGAN-GP further supports the result that PQCs have a stronger expressive power than classical neural networks \cite{pqc-expressive}. For our WGAN-GP architecture, the generator consists of roughly 1.46 million trainable parameters while our quantum generator consists of 5376 trainable parameters. Yet, our quantum generator is able to keep up with the classical generator. In comparison, a classical generator with a similar parameter count is unable to learn anything meaningful. In the future when the technology is sufficiently developed, it would be interesting to investigate how larger scale QGANs can compare to classical GANs in terms of performance.

\subsection{Binary FMNIST}
\label{section:binary-fmnist}

Next, we investigate whether the PQWGAN can learn from and generate more complex data, namely images from the classes of T-shirt and trousers from FMNIST. In this experiment, each of the sub-generators now has 11 layers to accommodate the increase in the data complexity. We keep the structure of having 7 data qubits and 1 ancilla qubits and observed that it worked well. 

The results in Figs. \ref{fig:image-gen}(e) and \ref{fig:image-gen}(f) show that again for both the classical and quantum case, the generators can successfully learn to generate images of T-shirts and trousers. However, the problems of samples being uncertain and fuzzy also persist in this case. With the increased complexity of the task, it is expected that these problems will be more apparent as the generator now has a harder time of generating a sharp image with more detail. Still, we observe that our PQWGAN is attempting to learn more subtle details, such as different shades and thickness of legs in the trousers of Fig. \ref{fig:image-gen}(e). This shows that there is potential for quantum generators to learn from more complex images in the future. 

Looking at the training curve for the PQWGAN in \ref{fig:image-gen}(h), the Wasserstein distance is not decreasing much and has a high variation as it gets updated. This suggests that the generator is nearing its capacity, and is failing to learn a representation that can capture all the variations in details of the samples it is generating due to its limited expressiveness. To rectify this problem, we could increase the number of layers in each sub-generator to increase the expressiveness of the generator. Furthermore, we could also increase the batch size of the learning process to obtain more stable gradients. However, both these methods are expensive to simulate as they increase the amount of resources required to simulate the circuits and are out of the scope for this work. 

\subsection{Triple MNIST}

After investigating the performance of our framework for image generation in two classes of a dataset, we now investigate whether our framework can be applied to generate more classes simultaneously. From Section \ref{section:binary-fmnist}, we observed that the generator struggles to properly learn the more complex and diverse features on FMNIST with its current size. Hence, we focused on the task of learning on three classes of MNIST. We selected the digits 0, 1 and 3 to learn as they have a distinct structure to them. In this case, we use 11 layers per sub-generator of 7 data qubits and 1 ancilla qubit.

Our results show that both the classical and quantum frameworks are able to generate images that correspond to the three digits. However, for the PQWGAN, there are now artifacts that persist in the same location of every output image. The reason for such simulation results is not fully known. However, it is empirically observed that varying certain parameters such as the number of data qubits in the quantum generator reduces the impacts of these artifacts. This will be further explored in Section \ref{section:effects}.

Ignoring these artifacts, both the classical and quantum frameworks are able to learn to generate images of 0, 1 and 3. However, there is an increased proportion of mixed images in PQWGAN when compared to Section \ref{section:binary-mnist}. This is expected as the dataset is now more complex, with fewer epochs. Furthermore, the Wasserstein distance as shown in Fig. \ref{fig:image-gen}(l) is plateauing, which suggests that the generator is nearing its capacity. Hence, we require more layers and a longer training process to better learn from triple MNIST.

\subsection{Walking in the latent space}

In classical GANs, interpolations between two points in the latent space have been used to demonstrate that the generator has learned to output a smooth mapping instead of memorising specific samples \cite{dcgan}. Here, we perform linear interpolation to visualise the mapping that the PQWGAN has learned. There exists more complex interpolation methods, such as spherical linear interpolation \cite{interpolation} for high dimensional latent spaces ($> 50$ dimensions). However, as the dimensions of our latent space is equal to the number of qubits in a sub-generator, it is of low dimensions ($< 10$ dimensions) due to resource constraints. Furthermore, since we are only interested in verifying that the generator has learned a smooth mapping for now, we use linear interpolation as it is more intuitive.

\begin{figure}[h]
    \centering
    \subfloat[0 to 1]{
      \includegraphics[width=0.95\linewidth]{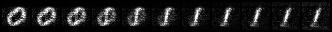}
     }
     \\
    \subfloat[7 to 9]{
        \includegraphics[width=0.95\linewidth]{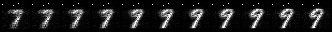}
    }
    \\
    \subfloat[1 to 7]{
      \includegraphics[width=0.95\linewidth]{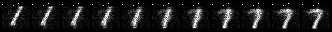}
    }
    \\
    \subfloat[T-shirt to trousers]{
      \includegraphics[width=0.95\linewidth]{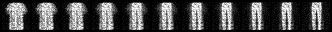}
    }
    \\
    \subfloat[sneakers to trousers]{
      \includegraphics[width=0.95\linewidth]{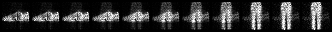}
    }
    \\
    \caption[PQWGAN walk]{\textbf{Linear interpolation on PQWGAN.} Performing linear interpolation between two points of the PQWGAN shows a smooth transition. This shows that our PQWGAN can learn a smooth mapping from the latent space to the output space.}
    \label{fig:walk}
\end{figure}

We pick two latent vectors corresponding to two well defined images and divide the straight line connecting them into ten equal segments. Then, we use the points that lie on the ends of these segments to generate outputs. The results are shown in Fig. \ref{fig:walk}. For all our experiments, including both the two and three class experiments, the generator is able to learn a smooth mapping for points in the latent space. This shows that our generator is indeed capable of learning a meaningful mapping from the latent space to the output space.

\section{Effects of generator parameters}
\label{section:effects}

To this end, we showed that our framework can be successfully applied to generate high resolution images of both MNIST and FMNIST, we now turn to investigate how the structure of the quantum generator will affect the quality of the images generated to provide guidance on how to pick the generator architecture. Since there are many different choices that one can make for the generator to generate an image, we choose a set of parameters that we believe has a significant impact on the learning process. Namely, we investigate how altering the number of patches, qubits and layers in the generator, the shape of the patches, and the choice of prior distribution affects the resulting image quality. In each experiment, we vary the parameter in question and fix all other parameters. The training curves obtained from these experiments can be found in Appendix \ref{appendix:effects-curves}.

\subsection{Number of patches}
\label{section:number-of-patches}

First, we investigate how the number of patches affects the quality of our generated samples. To focus on the effects of the number of patches, we use 1 ancilla qubit and the minimum number of data qubits required for the patch size in every sub-generator. Furthermore, we adjust the number of layers per sub-generator to keep a similar number of trainable parameters so that the expressiveness of the generator stays relatively constant. We pick the structure with 28 sub-generators and 10 layers as a baseline, then vary the number of layers with the number of batches accordingly. Our various generator structures are shown in Table \ref{tab:number-of-patches}, and we applied these generators to learn from MNIST 0/1 and FMNIST T-shirt/trousers.

\begin{table}[h]
    \centering
    \caption{Various generator structures used to investigate the effects of different number of patches.}
    \setlength{\tabcolsep}{3pt}
    \begin{tabular}{|M{55pt}|M{55pt}|M{55pt}|M{55pt}|}
        \hline
        Number of patches & Number of data qubits & Number of Layers & Total number
        of parameters \\
        \hline
        1 & 10 & 153 & 5049 \\
        2 & 9 & 84 & 5040 \\
        4 & 8 & 47 & 5076 \\
        7 & 7 & 30 & 5040 \\
        14 & 6 & 17 & 4998 \\ 
        28 & 5 & 10 & 5040 \\ 
        \hline
    \end{tabular}
    \label{tab:number-of-patches}
\end{table}

\begin{figure*}
    \begin{center}
        \includegraphics[width=\textwidth]{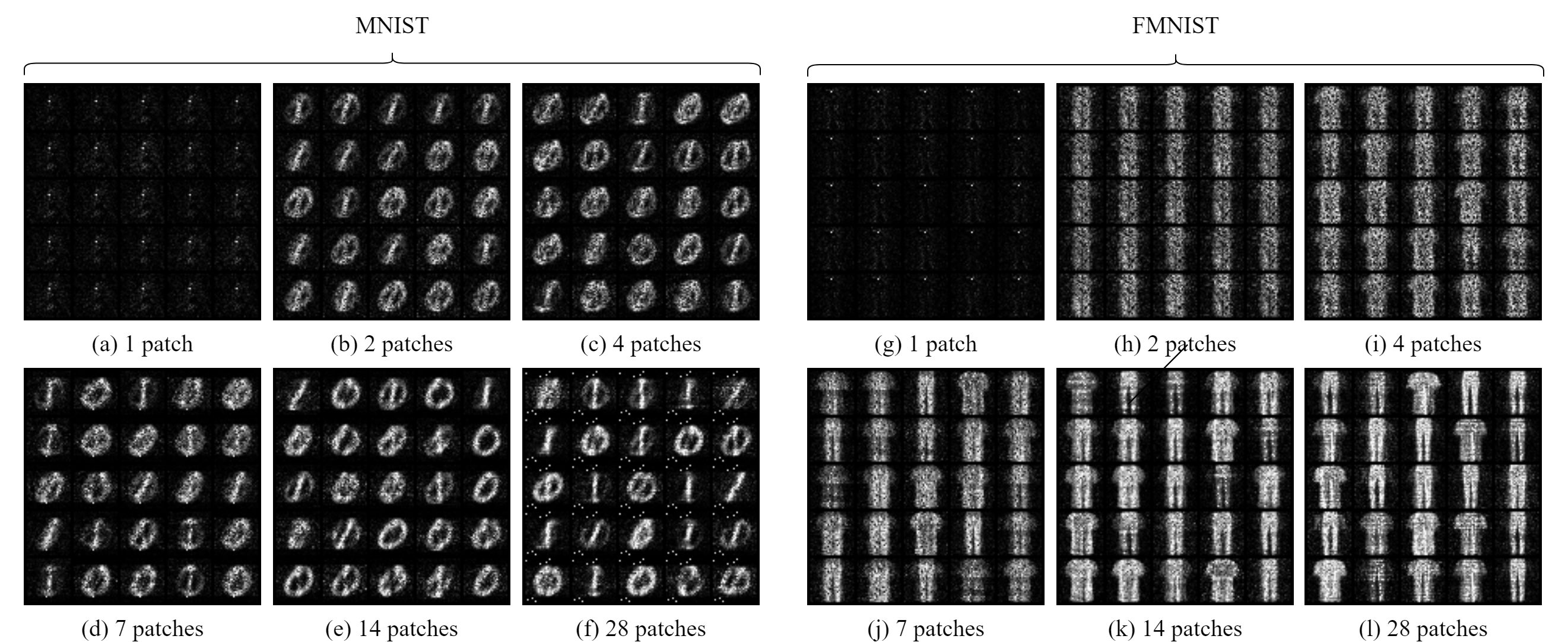}
        \caption{\label{fig:patches} \textbf{Effect of number of patches.} Varying the number of patches without changing other parameters will change the expressiveness of the generator by adding more trainable parameters. Hence, in our experiments we fixed the number of trainable parameters to be roughly constant and varied the number of data qubits and layers accordingly. The images show random outputs sampled from generators with varying number of patches trained on MNIST and FMNIST. In both cases, having a small number of patches led to poor learning. As we increase the number of patches, the generated images become sharper.}
    \end{center}
\end{figure*}

From Fig. \ref{fig:patches}, the results show that there is a significant effect in terms of the number of patches on the quality of the outputs. Starting with only 1 patch, the generator fails to output anything meaningful. Instead if we inspect intermediate outputs during the training process, the outputs oscillate between being dark and something that resembles a mode collapse. As we increase the number of patches, the images become increasingly sharper, and there are fewer mixed images.

The drop in the output quality as we decrease the number of patches is likely due to the effects of barren plateaus. It has been shown that for a hardware-efficient ansatz, training with a global cost function exhibits barren plateaus regardless of circuit depth \cite{cost-functions-barren-plateaus}. In our framework, the cost function is global as we are directly comparing the final state of the sub-generators to images, which can be thought of as generated from some arbitrary state. As we increase the number of patches, the effect of barren plateaus decreases due to having smaller quantum circuits in the sub-generators. Hence, the sub-generators are able to more effectively explore the Hilbert space, which allows it to quickly search for a mapping. This is evident from the training curves, where the Wasserstein distance converges to a lower value and also has less variance as we increase the number of patches in both experiments. 

On the other hand, artifacts do not exist when we have fewer patches for the MNIST experiments. Hence, using less patches may be useful for avoiding artifacts. However, the quality of the outputs decreases significantly as we decrease the number of patches. Hence, having more patches will in general corresponds to better output quality.

\subsection{Number of qubits}

Next, we investigate how the number of qubits affects the quality of our generated outputs. We focus on varying the number of data qubits used to generate a patch by increasing the number of data qubits while keeping the same number of patches and discarding an increasing number of pixels. Specifically, we focus on the architecture with 28 patches in our sub-generator, and have 5 to 8 data qubits while having 1 ancilla qubit. Again, to keep the expressiveness of the sub-generator similar, we keep the parameter count roughly the same by varying the number of layers accordingly. The list of generator configurations tested is shown in Table \ref{tab:qubits}. Since varying the data qubits may have an effect on the learning capabilities of the sub-generators, we apply these configurations to the more difficult tasks, namely MNIST 0/1/3 and FMNIST trousers/sneakers to obtain a more pronounced effect. 

\begin{table}[h]
    \centering
    \caption{Various generator structures used to investigate the effects of different number of
    data qubits.}
    \setlength{\tabcolsep}{3pt}
    \begin{tabular}{|M{70pt}|M{70pt}|M{70pt}|}
        \hline
        Number of data qubits & Number of Layers & Number of parameters \\
        \hline
        5 & 15 & 7560 \\ 
        6 & 13 & 7644 \\ 
        7 & 11 & 7392 \\ 
        8 & 10 & 7560 \\
        \hline
    \end{tabular}
    \label{tab:qubits}
\end{table}

\begin{figure*}
    \begin{center}
        \includegraphics[width=0.8\textwidth]{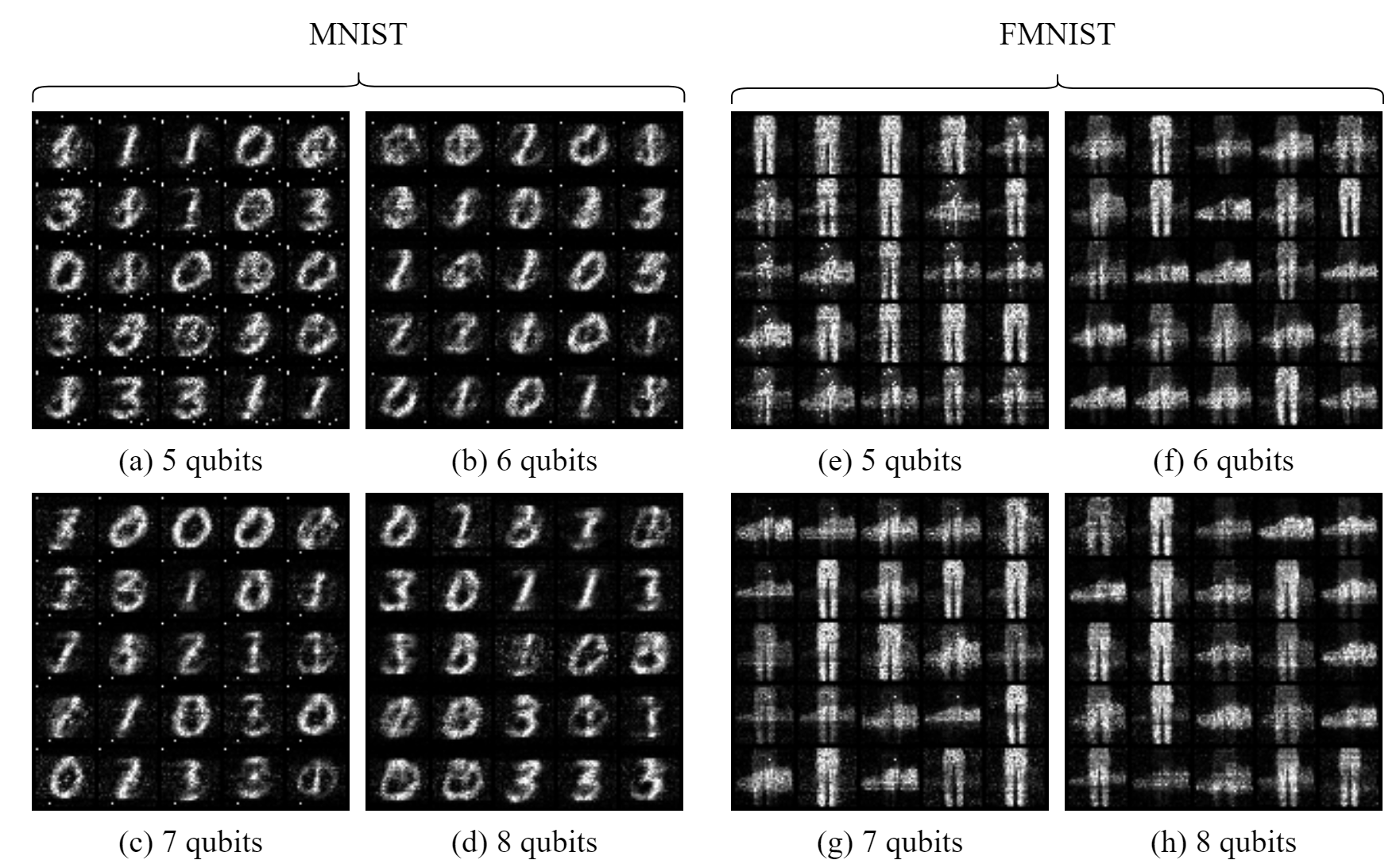}
        \caption{\label{fig:qubits} \textbf{Effect of number of qubits.} Varying the number of data qubits without varying the number of layers will change the expressiveness of the generator by adding more trainable parameters. Hence, in our experiments we fixed the number of trainable parameters to be roughly constant by using 28 patches while having a varying number of layers. The images show random outputs sampled from generators with varying number of data qubits trained on MNIST and FMNIST. In both cases as we increase the number of qubits, the outputs become sharper and less convoluted while also leading to less artifacts.}
    \end{center}
\end{figure*}

From Fig. \ref{fig:qubits}, the outputs from generators with extra data qubits are in general smoother and less convoluted, which is supported by the fact that the Wasserstein distance converges to a lower value when we increase the number of qubits. With more qubits, the generator has more flexibility as it can utilise the extra data qubits to manipulate the final amplitude into a higher quality image. However, there is a diminishing return in terms of output quality and Wasserstein distance as we continue to increase the number of qubits. From visually inspecting the outputs and training curve, there does not seem to be much benefit going past 6 qubits in terms of the image sharpness for these two tasks.

In addition to improving the output quality, we observe that having additional qubits helps in reducing the amount of artifacts present in the final outputs. Other than running our framework with more data qubits on these two tasks, we also tried having more data qubits for the experiments in Section \ref{section:image-generation} when trying to achieve better outputs. We observed a similar trend, where in general having more data qubits also reduced the amount of artifacts present in the learned outputs. To investigate this, we inspected the outputs of the quantum circuits of the sub-generators for our two experiments. Since we are running simulations, we can obtain precisely the quantum state produced at the end of the quantum circuits. We observed that there is a large difference between the sum of the probabilities of the basis states that are used and discarded. In both experiments for the 5 qubit case, the sum of the probabilities of the used basis states were very often greater than 0.9, while for the 8 qubit case it was very often less than 0.1. This suggests that the presence of artifacts is due to the lack of excess qubits for the sub-generators to iron out the imperfections in a highly entangled output state.

\subsection{Number of layers}

We now turn to investigate the effect of having different number of sub-generator layers. Unlike previous experiments, varying the number of layers will directly change the number of parameters, which affects the expressiveness of the circuit. As we add more layers, the quantum circuit can learn more complex transformations of the input quantum state, which should allow the generator to learn a more complex distribution. Hence, to observe a pronounced effect, we again apply it to the more complex tasks of generating MNIST 0/1/3 and FMNIST trousers/sneakers. To be consistent with previous experiments and to conserve resources, we use 28 patches with 5 data qubits and 1 ancilla qubit for each generator. We experimented with having 5, 10, 15, 20 and 25 layers of parameterised gates.

\begin{figure}
    \centering
    \subfloat[]{
      \includegraphics[width=0.8\linewidth]{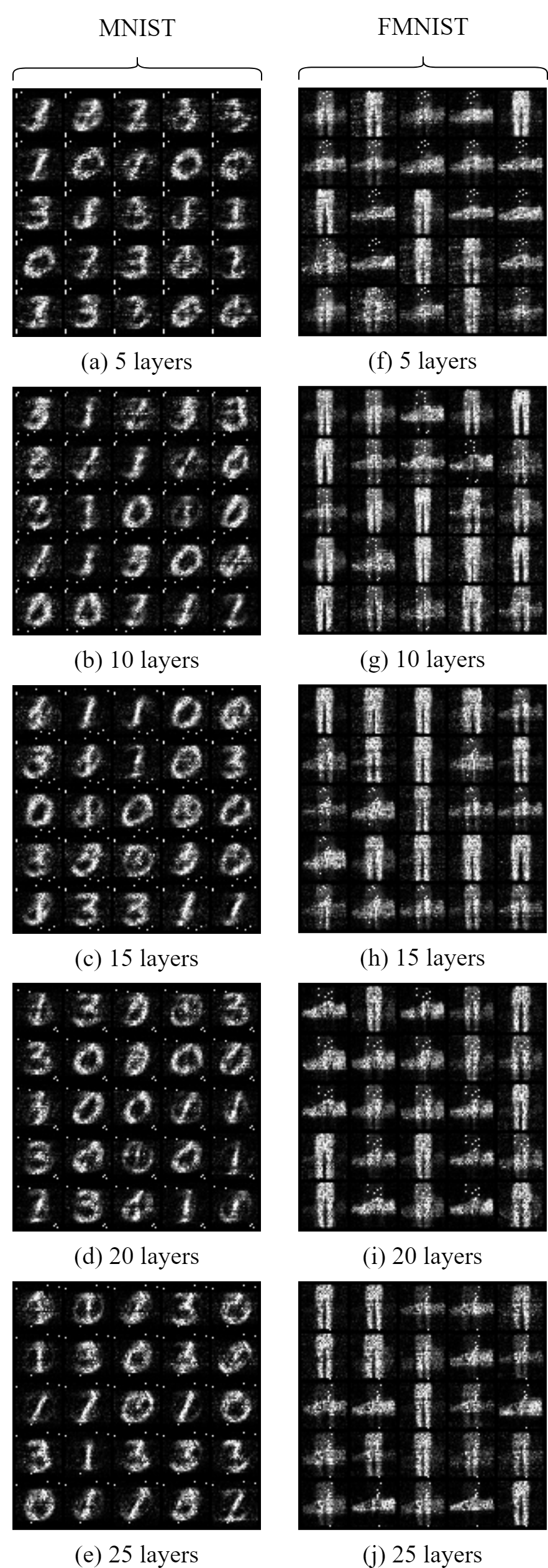}
     }
     \caption{\label{fig:layers} \textbf{Effect of number of layers.} In this experiment, we are interested in having a more expressive generator. Hence, we fixed the number of patches and data qubits to be 28 and 5 respectively. The images show random outputs sampled from generators with varying number of layers trained on MNIST and FMNIST. In both cases, increasing the number of layers improves the sharpness of the outputs. However, there is a diminishing return as we keep increasing the number of layers. }
\end{figure}

From Fig. \ref{fig:layers}, the quality of the outputs increases with more layers and there are less mixed images. Furthermore, the Wasserstein distance converges to a lower value albeit having greater variance. However, the gain is marginal as we keep increasing the number of layers, and there is not a big improvement as we go past 15 layers. The marginal increase in output quality and the larger variance in Wasserstein distance estimates can be explained by the expressibility of PQCs \cite{expressibility}. As we add more layers to our sub-generators, it can represent a larger set of unitaries $\mathbb{U}$. Hence, as we add more layers to our generator, it becomes increasingly likely that the generator is complete, and that $\mathbb{U}$ contains an acceptable solution. However, the larger set of unitaries also means that the optimisation process has to search through a larger space for the solution. Hence, it is likely that the optimisation process has to navigate a more complex parameter space, which leads to a higher variance in the Wasserstein distance.

On the other hand, the fact that the Wasserstein distance starts to plateau in all configurations suggests that our choice of ansatz may not be the most suitable for the image generation task. Our choice of the PQC structure is chosen such that it can represent any hardware-efficient ansatz of repeating single qubit rotations followed by CNOT gates. However, there are no theoretical motivations as to why we should use a hardware-efficient ansatz instead of other types of ansatz for image generation. Problem-inspired ansatz such as the quantum alternating operator ansatz \cite{problem-inspired-ansatz} have been shown to be useful in more efficiently implementing solutions to combinatorial optimisation problems. Thus, it is possible that there are other types of ansatz that are more suitable for image generation.

\subsection{Shape of Patches}

Next, we investigate how the shape of the generated patches affects the final outputs. In previous experiments, the data points generated by a patch always fit entirely in one row and are then stacked vertically. In this experiment, we considered having $7 \times 4$ patches and stacking them across horizontally. We are now more focused on whether the generator is able to generate images with a different layout of patches, and so we apply it to the easier tasks of generating MNIST 0/1 and FMNIST T-shirts/trousers. Again, we use 28 patches of 10 layers with 5 data qubits and 1 ancilla qubit for consistency.

\begin{figure*}
    \begin{center}
        \includegraphics[width=0.8\textwidth]{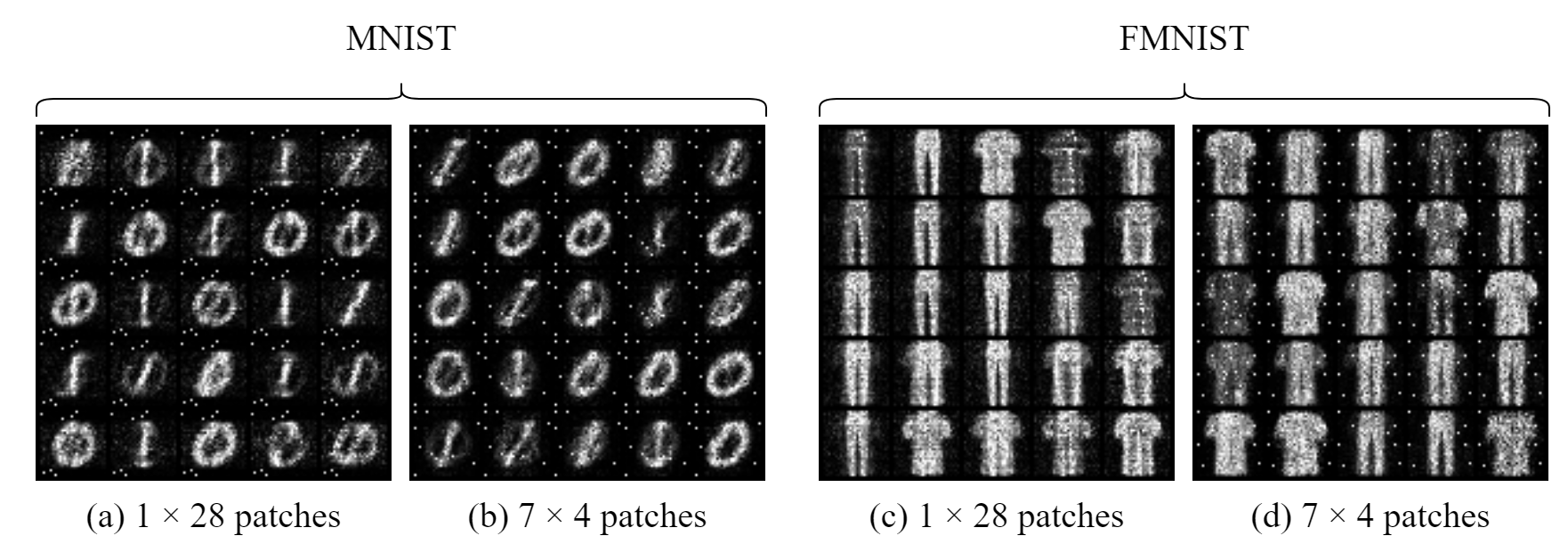}
        \caption{\label{fig:patch-shape} \textbf{Effect of shape of patches.} In this experiment, we fixed the number of patches, layers and data qubits to be 28, 10, and 5 respectively and varied the shape of the patches. The images show random outputs sampled from generators with either $1 \times 28$ or $7 \times 4$ patches trained on MNIST and FMNIST. In both cases, having different patches shapes do not affect the quality of the outputs. However, should artifacts exist, having different patch shapes can lead to artifacts appearing in more obvious locations.}
    \end{center}
\end{figure*}

From Fig. \ref{fig:patch-shape}, the shape of the patches does not have a big effect on the final output quality. This is also supported by the training curves, where the evolution of Wasserstein distance is similar in both patch shapes. However, there is a noticeable gap in the Wasserstein distance estimates, which is due to the location of artifacts in the $7 \times 4$ patch images. Due to the way the patches are positioned, any artifacts generated using the $7 \times 4$ patches will appear in center of the final image. Hence, the critic has a very easy time of spotting imperfections that exist, leading to a larger Wasserstein distance estimate.

Ignoring the artifacts, both patch shapes were still able to learn to produce images of the corresponding objects. Given that we already know it works for $1 \times 28$ patches, we would expect the generator to be able to successfully learn regardless of the intended shape of its outputs. Initially, we chose having $1 \times 28$ patches as it was easier to implement, and did not take into account any of the structure of the images. Since it was able to learn successfully, it showed that the optimisation process can successfully guide the patches in producing what it needs for that region. Hence, changing the patch shape would not affect whether the generator can learn to output images. Having said that, changing the patch shape may be useful for more complex datasets with more structure to the images. For example, we can imagine for CIFAR-10 \cite{cifar} images of natural scenes, it might be beneficial to have patches that corresponds to certain segments of the photo, such as the background, foreground and objects within.

\subsection{Prior distribution}

Finally, we investigate the effect of different prior distributions on the outputs. In classical GANs, the prior distribution is usually chosen to be a Gaussian due to its nice mathematical properties and empirical performance. However in our preliminary testing, we observed that the generator only learns to output meaningful samples on some of the inputs from a Gaussian prior. Hence, we opted to use a uniform prior instead. Here, we make a more detailed comparison of the PQWGAN trained on a Gaussian prior and uniform prior. As before, the uniform prior is chosen to be from $U_{[0,1)}$, while the Gaussian prior is chosen to be the standard normal $\mathcal{N}(0,1)$. We applied it to the task of generating MNIST 0/1/3 and FMNIST trousers and sneakers using 28 patches of 5 data qubits, 1 ancilla qubit and 15 layers.

\begin{figure*}
    \begin{center}
        \includegraphics[width=0.8\textwidth]{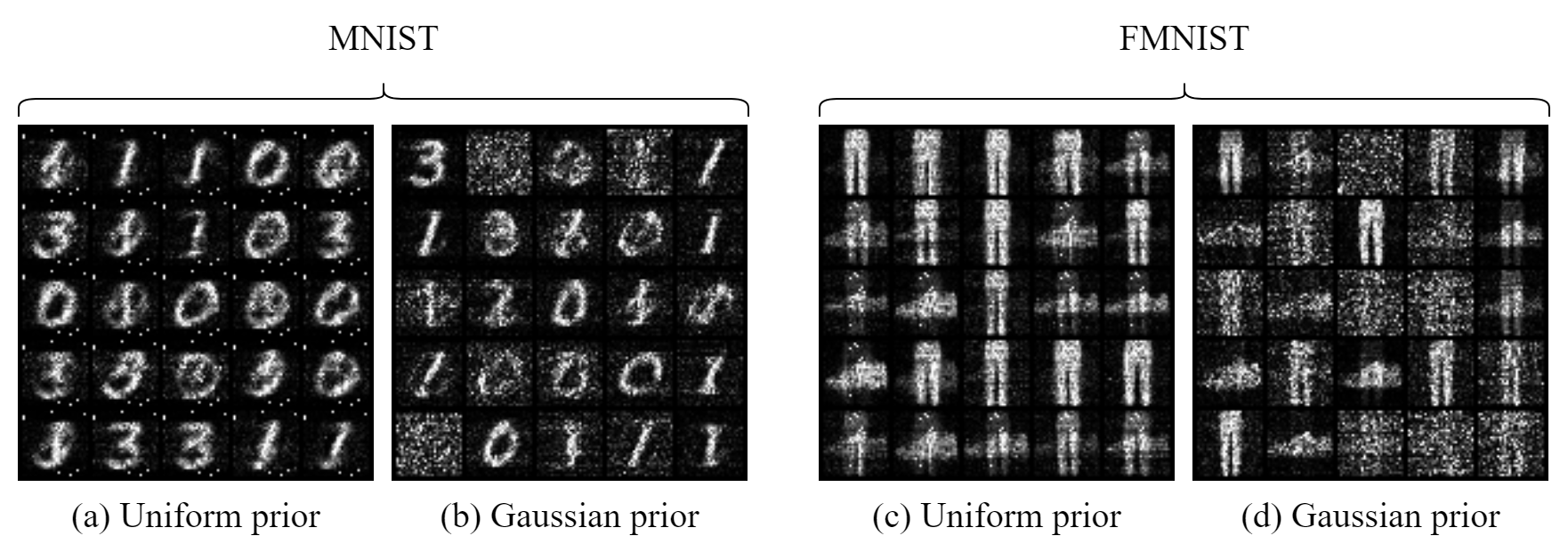}
        \caption{\label{fig:prior} \textbf{Effect of prior distribution.} In this experiment, we fixed the number of patches, layers and data qubits to be 28, 15, and 5 respectively and varied the choice of prior distribution. The images show random outputs sampled from generators with a uniform or Gaussian prior trained on MNIST and FMNIST. In both cases, the uniform prior allows the generator to learn a more complete mapping for every latent vector, while the Gaussian prior allows the generator to produce sharper images for those that are well formed.}
    \end{center}
\end{figure*}

From Fig. \ref{fig:prior}, in both cases the generator with the Gaussian prior has not learned to fully map the latent space to the output space while the uniform prior has, which is due to the small number of training iterations done coupled with the low-dimensional latent space. Since the latent vectors sampled from the Gaussian distribution are concentrated around the mean, the small number of training iterations means that the generator has potentially not fully explored the latent spaces far from the mean. For the uniform prior, this does not happen as the latent vectors are sampled with equal probability. Hence, in our case the generator will likely have learned a more complete mapping of the latent space when using a uniform distribution.

On the other hand, the well formed samples generated from a Gaussian prior are arguably of better quality than compared to the uniform prior. Especially for the MNIST images such as row 1 column 1, row 2 column 1 and row 5 column 2 of Fig. \ref{fig:prior}(b), the images appear sharper and have less noise compared to those in Fig. \ref{fig:prior}(a). Furthermore, the outputs from the Gaussian prior are free from artifacts. Since the generator is frequently trained on latent vectors centered around the mean, the outputs generated from this region of the latent space will be of better quality. As the training progresses using a Gaussian prior, we observed that there are less instances of white noise generated by the generator, which supports the argument of the white noise coming from an unexplored region of the latent space. In the future when there are more resources to run larger scale training, the PQWGAN trained on a Gaussian prior may generate better outputs than using a uniform prior while avoiding the problem of artifacts.

\subsection{Summary of effects of generator parameters}

In this section, we have investigated the effects of various parameters on our quantum generator. Here, we summarise some general observations on the impact of different parameter choices on the output quality. First, in terms of number of patches to use, we observe that having more patches in general corresponds to better output quality. In our experiments, we were unable to learn anything meaningful with 1 patch, and the output quality improved as we increased the patches. Next, as we increase the number of data qubits, the outputs were sharper and less artifacts were observed. Third, as we increased the number of layers in a sub-generator, the images are visibly sharper when we go from 5 to 10 and to 15 layers but are less noticeable from 15 layers onwards. Fourth, the shape of the patches do not have much impact on the quality and general sharpness of the outputs, but having square patches will lead to more noticeable artifacts if they exists. Finally, using a uniform prior allows the generator to quickly explore the whole latent space while the Gaussian prior is slower and leads to more white noise in our small scale experiments. However, the well-formed outputs generated by the Gaussian prior are arguably of higher quality and do not suffer from artifacts.

\section{Conclusions and future directions}
\label{section:conclusion}

Classical GANs have seen great success in image generation while QGANs are still far away from that level. However, with the promise that quantum computing brings, QGANs are an exciting area of research. In this paper, we proposed our novel PQWGAN framework for image generation and empirically evaluated the effects of varying its different parameters. The PQWGAN is composed of a quantum generator and classical critic, and is inspired by the patch strategy QGAN and the training algorithm of WGAN-GP. We applied our PQWGAN to learn on subsets of $28 \times 28$ pixels images of MNIST and FMNIST. Specifically, we successfully learned the two classes subsets of MNIST 0/1 and FMNIST T-shirts/trousers, and the three class subset of MNIST 0/1/3. We compared these results to a WGAN-GP and found that our PQWGAN was able to achieve comparable results despite having 3 orders of magnitude less trainable parameters in the generator. To the best of our knowledge, this is the first time that a QGAN with a quantum generator has demonstrated that it can successfully learn multimodal image data of this resolution on these standard datasets. Finally, we investigated how varying different parameters has an effect on the output image quality. We found that as a general rule of thumb, having more patches, data qubits, layers is beneficial, while the shape of the patches did not matter too much. We also found that using a uniform prior is beneficial in the near term, but may be outperformed by a Gaussian prior in the future.

We suggest four future directions to explore. First, our work here is based solely on noiseless simulations. Future work could look to perform simulations with different noise models to investigate the effects of noise on the viability of the framework. Furthermore, as quantum hardware continues to improve, physical quantum computers with more qubits and lower error rates are expected to arrive in the near future. Since our framework is capable on running on NISQ devices, future work could also look to implement our framework on physical devices. The patch strategy QGAN \cite{p-qgan} successfully learned to generate simple $8 \times 8$ pixels images on a 12-qubit superconducting computer. With 100+ qubit devices currently available and 1000+ qubit devices projected to arrive in the near future, it would be interesting to observe how our framework performs on a real quantum computer. Second, we could try more complex images such as natural scenes with CIFAR-10 \cite{cifar} or human faces with CelebA \cite{celeba} in future work to see how our framework holds up in the presence of color and irregular image structure. Third, we noted that our choice of the quantum circuit for the generator lacked theoretical motivations. Like how convolutions have allowed deep learning for images to leap forward, there may be some operations that a quantum circuit can do that are especially helpful for images. Future work could look to experimenting with different ansatz to discover image manipulation techniques more suited for QML. Finally, our critic in this work is implemented using classical neural networks. As the field of QML develops, having a quantum critic can be beneficial due to the promises that QML brings. Combining the last two directions, our framework has the potential to be the foundation of new methods to push forward what is currently capable of QGANs, and work towards closing the gap in what is possible with classical and quantum machine learning methods.

\appendix

\section{Details of the PQWGAN framework}
\label{section:details}

\subsection{Comparison of our work to existing QGANs}

A summary of the related works on image generation with QGANs mentioned in Section \ref{section:related} can be found in Table \ref{tab:comparison}.

\begin{table}[ht]
    \centering
    \caption{A comparison of our work to existing QGANs for image generation.}
    \setlength{\tabcolsep}{3pt}
    \begin{tabular}{|M{55pt}|M{50pt}|M{45pt}|M{40pt}|M{40pt}|}
        \hline
        Paper & Dataset(s) & \parbox[m]{45pt}{Output size \vspace*{1pt}} & \parbox[m]{40pt}{Maximum number of classes \vspace*{1pt}} & Notes \\
        \hline
        This work & \parbox[m]{50pt}{MNIST, FMNIST \vspace*{1pt}} & \parbox[m]{45pt}{$28 \times 28$ pixels \vspace*{1pt}} & 3 & \parbox[m]{40pt}{Does not require pre/post-processing \vspace*{1pt}} \\
        \rule{0pt}{4ex}
        \parbox[m]{55pt}{Experimental Quantum Generative Adversarial Networks for Image Generation \cite{p-qgan} \vspace*{1pt}} & \parbox[m]{50pt}{Handwritten digits \vspace*{1pt}} & $8 \times 8$ pixels & 2 & \parbox[m]{40pt}{Low image quality \vspace*{1pt}} \\
        \rule{0pt}{4ex}
        \parbox[m]{55pt}{QuGAN: A Quantum State Fidelity based Generative Adversarial Network \cite{QuGAN} \vspace*{1pt}} & MNIST & \parbox[m]{45pt}{4 dimensions \vspace*{1pt}} & 3 & \parbox[m]{40pt}{Uses PCA to compress images \vspace*{1pt}} \\
        \rule{0pt}{4ex}
        \parbox[m]{55pt}{IQGAN: Robust Quantum Generative Adversarial Network for Image Synthesis On NISQ Devices \cite{iqgan} \vspace*{1pt}} & MNIST & \parbox[m]{45pt}{16 dimensions \vspace*{1pt}} & 3 & \parbox[m]{40pt}{Uses PCA to compress images \vspace*{1pt}} \\
        \hline
    \end{tabular}
    \label{tab:comparison}
\end{table}

\subsection{Inner workings of the quantum generator}
\label{section:generator-details}

The inner workings of our quantum generator is very similar to that of \cite{p-qgan}. The generator first takes in a $N$-dimensional latent vector $\boldsymbol{z} = (z_1, z_2, ..., z_N)$ from some distribution $p_{\boldsymbol{z}}$ (eg. uniform, Gaussian). The latent vector is then encoded in each sub-generator using a layer of $R_Y$ rotations parameterised by the components of $\boldsymbol{z}$. So, starting in the $\ket{0}^{\otimes n}$ state, we obtain the latent state $\ket{\boldsymbol{z}}$ by applying the encoding circuit 
\begin{equation*}
    \ket{\boldsymbol{z}} =  R_Y^1(z_1) R_Y^2(z_2) ... R_Y^N(z_N) \ket{0} ^ {\otimes N}
\end{equation*}
where $R_Y^i(z_i)$ is the $R_Y$ gate applied to the $i$-th qubit with the rotation angle $z_i$. Then, the latent state passes through the $L$ layers of the hardware-efficient ansatz structure \cite{hardware-efficient}, each consisting of parameterised arbitrary rotations $R(\phi, \theta, \omega)$ followed by CNOT gates on each adjacent qubit to generate entanglement. The $R(\phi, \theta, \omega)$ gate can be expressed as 
\begin{multline*}
    R(\phi, \theta, \omega) = R_Z(\omega) R_Y(\theta) R_Z(\phi) = \\
    \begin{bmatrix} e^{-i(\phi + \omega)/2} \cos(\theta/2) & e^{-i(\phi - \omega)/2} \sin(\theta/2) 
    \\ e^{-i(\phi - \omega)/2} \sin(\theta/2) & e^{-i(\phi + \omega)/2} \cos(\theta/2) \end{bmatrix}
\end{multline*}
This gate was chosen as it can represent any single qubit rotation that we want up to a phase shift. Furthermore, it can be easily decomposed as a series of $ZYZ$ gates, which can be implemented on a real device. The $L$ parameterised layers act essentially as one big unitary operation $\mathcal{U}_L(\boldsymbol{\phi_i, \theta_i, \omega_i})$ that performs a linear transformation on the state $\ket{\boldsymbol{z}}$, and the resulting quantum state generated by the $i$-th sub-generator is
\begin{equation*}
    \ket{\psi_{G_i}} =  \mathcal{U}_L(\boldsymbol{\phi_i,\theta_i,\omega_i}) \ket{\boldsymbol{z}}
\end{equation*}
The success of deep learning methods such as neural networks lies in its ability to learn non-linear transformations of its input. To introduce non-linearity into the sub-generators, we make a partial measurement $M$ on the ancilla qubits, then trace out the ancilla qubits to obtain the resulting state of the data qubits. Since we will be making projector measurements, the state of the data qubits $\ket{\psi_{D}}$ after tracing out the ancilla qubits will be
\begin{equation*}
    \ket{\psi_{D}} = \text{Tr}_A \left(\frac{M \otimes \mathbb{I} \ket{\psi_{G_i}} \bra{\psi_{G_i}}}
    {\braket{\psi_{G_i} | M \otimes \mathbb{I} | \psi_{G_i}}} \right)
\end{equation*}
In our case, we pick the partial measurement to be $M = (\ket{0} \bra{0})^{\otimes A}$ for simplicity. Hence the final state of the data qubits will be 
\begin{equation*}
    \rho_D = \text{Tr}_A \left(\frac{(\ket{0} \bra{0})^{\otimes A} \otimes \mathbb{I} \ket{\psi_{G_i}} \bra{\psi_{G_i}}}
    {\braket{\psi_{G_i} | (\ket{0} \bra{0})^{\otimes A} \otimes \mathbb{I} | \psi_{G_i}}} \right)
\end{equation*}
The state now depends on $\ket{\psi_{G_i}}$ in both the denominator and the numerator, which is in turn dependent on $\ket{\boldsymbol{z}}$. Hence, the state is a non-linear transformation of $\ket{z}$. We then measure the probability of each computational basis state of the data qubits to obtain the sub-generator output given by
\begin{equation*}
    G_i(\boldsymbol{z}) = [p(0), p(1), ..., p(2^{D-1})]
\end{equation*}
We would like each element of the generator output to have values between $[0,1]$ to be interpreted as pixel values. Although it is possible to interpret the probabilities as pixel values directly, it would be problematic due to the normalisation constraint, which would not give us the desired pixel values. Hence, we apply post processing by taking 
\begin{equation*}
    G'_i(\boldsymbol{z}) = \frac{G_i(\boldsymbol{z})}{\max (G_i(\boldsymbol{z}))}
\end{equation*}
in order to obtain valid pixel values. Since the size of the outputs from the quantum circuit are power of 2s, we only keep the first $\frac{HW}{P}$ pixels to create a patch with the correct dimensions. Finally, the output from all the sub-generators are stacked together to form an image of size $H \times W$
\begin{equation*}
    G(\boldsymbol{z}) = [G'_1(\boldsymbol{z}),..., G'_P(\boldsymbol{z})]^T
\end{equation*}
It is possible to use other more complex non-linear transformations from classical machine learning, such as passing the output through different activation functions as discussed in \cite{p-qgan}. However, due to time constraints we did not investigate this further.

\subsection{Potentials of a quantum critic}

A quantum critic is an interesting research avenue in the future. In classical GAN training, the discriminator and generator architectures are usually chosen such that one does not significantly overpower the other \cite{gan-tutorial}. Although it is unclear whether this is a desirable property to have in GANs, having a rough idea of the expressive power of the GAN components can be used to help stabilise the training process. However, the relationship between the expressibility of classical neural networks and PQCs is unclear. Hence, having a quantum critic will be advantageous in this case as it can give a rough estimate of the expressiveness of the generator and critic. 

In practice, a quantum critic may be constructed in the same way as a sub-generator with PQCs. Instead of measuring all qubits, we can measure one qubit at the end of the circuit to obtain a value, such as the $Z$ expectation value. Then, to obtain an unbounded real valued output to serve as an estimate of the Wasserstein distance, we could pass the output value through a $\tan$ function, similar to an activation in a classical neuron. However, due to the limited time and scope of this work, we were not concerned with this.

\subsection{Discussions of the training process}
Ref. \cite{wgan} proved that for WGAN, the optimisation process is principled when the critic and generator are constructed with neural networks. Here, we argue that this is also the case for our form of PQCs. We rely on the following assumption and theorem proved in \cite{wgan}.

\begin{assumption}
    \label{as:WGAN}
    Let $g: \mathcal{Z} \times \mathbb{R}^d \to \mathcal{X}$ be locally
    Lipschitz between finite dimensional vector spaces. Denote $g_{\theta}(z)$
    as the result of evaluating $g$ with parameters $\theta$ at $z$. $g$
    satisfies the assumption for some probability distribution $p$ over
    $\mathcal{Z}$ if there exists local Lipschitz constants $L(z, \theta)$ such
    that 
    \begin{equation*}
        \mathbb{E}_{z \sim p}[L(z, \theta)] < + \infty
    \end{equation*}
\end{assumption}
\begin{theorem}
    \label{thm:WGAN}
    Let $\mathbb{P}_r$ be some distribution. Let $\mathbb{P}_{\theta}$ be the
    distribution generated by $g_{\theta}(z)$ where $g$ is some function
    satisfying assumption \ref{as:WGAN} and $z$ is some random variable with
    density $p(z)$. There exists a solution $f: \mathcal{X} \to \mathbb{R}$ to
    \begin{equation}
        \label{eq:distance-estimate}
        \max_{||f||_L \leq 1} \mathbb{E}_{x \sim \mathbb{P}_r} [f(x)] - \mathbb{E}_{x \sim \mathbb{P}_{\theta}} [f(x)]
    \end{equation}
    and
    \begin{equation*}
        \nabla_{\theta} W(\mathbb{P}_r, \mathbb{P}_{\theta}) = - \mathbb{E}_{z \sim p(z)}[\nabla_{\theta}f(g_{\theta}(z))]
    \end{equation*}
    when both terms are well defined.
\end{theorem}

Intuitively, Theorem \ref{thm:WGAN} tells us that it is possible for the generator to learn the target distribution under the WGAN objective as defined in (\ref{eq:WGAN}) using the min-max method in GANs. The inner maximisation corresponds to the Wasserstein distance reformulated under the Kantorovich-Rubinstein duality. By searching for a function $f$ from the family  of 1-Lipschitz functions that maximises the difference in expectations in (\ref{eq:distance-estimate}), we obtain the Wasserstein distance between our target distribution and our generator distribution. We would like to minimise the Wasserstein distance between the target distribution and generator distribution, and hence we can use the usual gradient descent of on the Wasserstein distance. This corresponds to the outer minimisation in the WGAN objective. With the gradient penalty term in WGAN-GP, the optimal critic is shown to have unit norm for straight lines connecting samples from $\mathbb{P}_r$ and $\mathbb{P}_{\theta}$. Hence, if expected value of the norm of the gradient deviates from 1, the critic will be penalised and will unlikely be the one that maximises (\ref{eq:distance-estimate}). This preserves the nice properties of Theorem \ref{thm:WGAN} and is empirically observed to support this claim. So, to show that the training of PQWGAN is also principled, it suffices to show that the quantum generator satisfies Assumption \ref{as:WGAN}. Using the theorem proved in \cite{pqc-lipschitz} on the Lipschitz continuity of PQCs, we
argue that our generator satisfies this assumption.

\begin{theorem}
    \label{thm:PQC_Lipschitz}
    Given a function $f:[0,2\pi]^M \to \mathbb{R}$ of the form
    $f(\boldsymbol{\theta}) = \bra{\psi} U^{\dagger}(\boldsymbol{\theta}) O
    U(\boldsymbol{\theta}) \ket{\psi}$, where $\ket{\psi} \in \mathbb{C}^n$ is
    some arbitrary state for a finite $n$, $U(\boldsymbol{\theta})$ is a quantum
    circuit parameterised by $\boldsymbol{\theta}$ consisting of an arbitrary
    number of fixed gates, and $M$ parameterised gates $U_i(\theta_i) =
    \exp(-i(\theta_i + c_i)H_i)$ for some constant $c_i$ and Hermitian $H_i$.
    Then, for any observable $O$ and any set of Hermitian operators,
    $f(\boldsymbol{\theta})$ is L-Lipschitz with 
    \begin{equation*}
        L = \sqrt{M} \left[\max_i \left(\sup_{\boldsymbol{\theta}} \left( \left|
        \frac{\partial f(\boldsymbol{\theta})}{\partial \theta_i}\right| \right) \right) \right]
    \end{equation*}
\end{theorem}

Since each sub-generator is a PQC of the form stated in Theorem \ref{thm:PQC_Lipschitz}, the sub-generators are Lipschitz continuous. Then, to obtain our final output, we apply a series of linear transformations to the sub-generator outputs, which preserves Lipschitz continuity. Hence, our full generator satisfies assumption 1 and the optimisation process is principled.

\section{Additional image generation experiments}

\subsection{Complex binary FMNIST}

\begin{figure}
    \centering
    \subfloat[PQWGAN]{
      \includegraphics[width=0.3\linewidth]{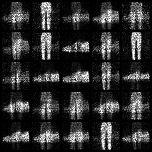}
     }
    \subfloat[WGAN-GP]{
        \includegraphics[width=0.3\linewidth]{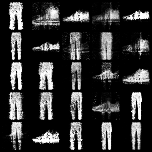}
    }
    \subfloat[Real samples]{
      \includegraphics[width=0.3\linewidth]{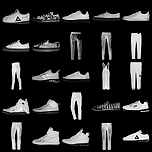}
     }
    \\
    \subfloat[]{
        \includegraphics[width=0.8\linewidth]{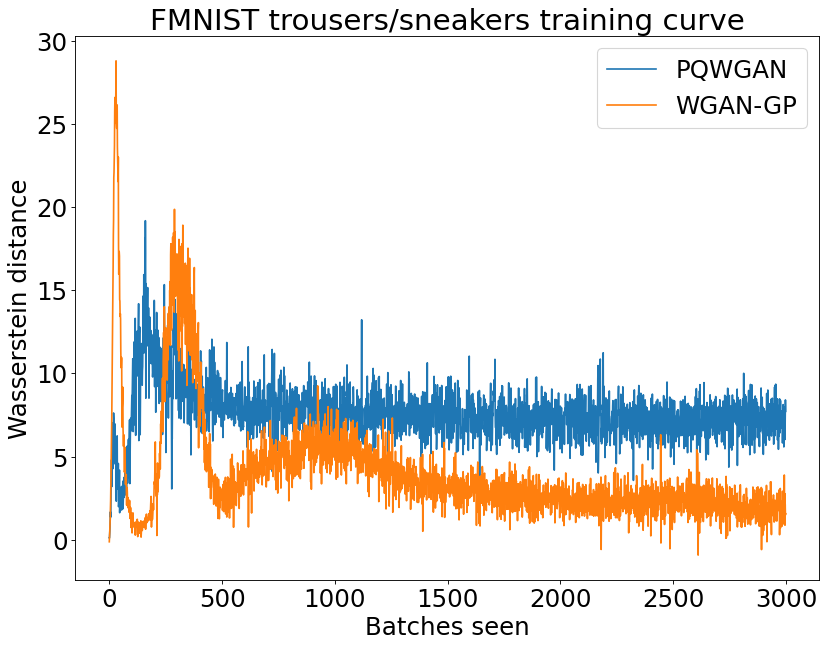}}
    \caption[PQWGAN on FMNIST 1/7]{Samples and training curves of classical and
    quantum architecture on FMNIST trousers/sneakers. These samples are generated randomly from
    (a) PQWGAN, (b) WGAN-GP and (c) the dataset. In (d) we show the tracked Wasserstein 
    distance of PQWGAN and WGAN-GP during training.}
    \label{fig:FMNIST-17}
\end{figure}

For the case of FMNIST T-shirt/trousers in the main text, the generator is having an easier time as since many of the pixels of the T-shirt and trousers are overlapping. So, the difference between them are mostly around the sleeves of the T-shirt and between the legs of the trousers. Hence, we also investigated whether the our framework can learn distinctly different patterns in the form of trousers and sneakers. In this experiment, each sub-generator consists of 13 layers of 6 data qubits and 1 ancilla qubits, and thus the latent space will have 7 dimensions. The results are shown in Fig. \ref{fig:FMNIST-17}.

Compared to the previous results, the PQWGAN is struggling more to converge to a set of parameters that can generate a comprehensive mapping of the latent space in this case. This is evident when we look at the training curve, where the Wasserstein distance of the PQWGAN converges and plateaus at a high value even if we continue to train it. Compared to the binary cases in the main text, there is a higher proportion of mixed samples of trousers and sneakers. This behaviour is expected since the dataset is both more complex and has a greater difference between the two classes that we are trying to learn. Again, this points to the problem of having a limited expressiveness of the generator, and hence being unable to capture the details of the dataset we are learning from. However, the PQWGAN is clearly still able to learn something meaningful, as it not only is able to generate images of trousers and sneakers, but also there exists some different details within the same class such as different shapes of sneakers in row 3 column 5 and in row 5 column 5 of Fig. \ref{fig:FMNIST-17}(a). Hence, it is reasonable to expect that in the future when more resources are available, we can have larger models and training processes that can mitigate this problem.

\subsection{Triple MNIST}

\begin{figure}
    \centering
    \subfloat[PQWGAN]{
      \includegraphics[width=0.3\linewidth]{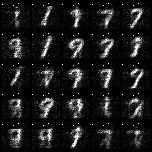}
     }
    \subfloat[WGAN-GP]{
        \includegraphics[width=0.3\linewidth]{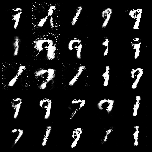}
    }
    \subfloat[Real samples]{
      \includegraphics[width=0.3\linewidth]{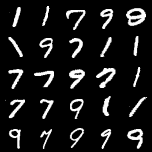}
     }
    \\
    \subfloat[]{
        \includegraphics[width=0.8\linewidth]{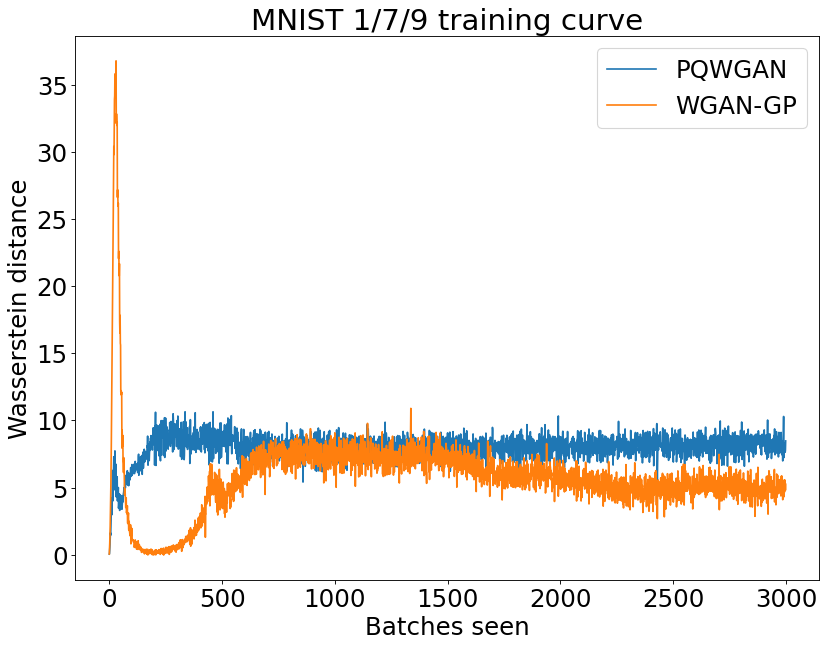}}
    \caption[PQWGAN on MNIST 1/7/9]{Samples and training curves of classical and
    quantum architecture on MNIST 1/7/9. These samples are generated randomly from
    (a) PQWGAN, (b) WGAN-GP and (c) the dataset. In (d) we show the tracked Wasserstein 
    distance of PQWGAN and WGAN-GP during training.}
    \label{fig:MNIST-179}
\end{figure}

We also applied our PQWGAN to the simpler task of learning the digits 1, 7 and 9. Compared to the MNIST 0/1/3 in main text, the digits are similar in the sense that they are all have some form of a vertical stroke, and hence should be easier to generate. In this experiment, we use 11 layers per sub-generator and 7 data qubits. The results are shown in Fig. \ref{fig:MNIST-179}.

 In this experiment, both the classical and quantum framework are able to learn to output images that corresponds to the three digits. If we ignore the existence of the artifacts, the PQWGAN samples have a comparable quality to the WGAN-GP samples. Some of the samples generated from both these frameworks closely resemble the real samples, while others are still noisy. Furthermore, from the training curve, the PQWGAN achieves a similar Wasserstein distance compared to the WGAN-GP in the early stages of training. However, the WGAN-GP slowly converges to a lower score while it seems that the PQWGAN is starting to plateau, which can be attributed to the generator being not expressive enough for the PQWGAN.

\section{Training curves of parameter experiments}
\label{appendix:effects-curves}

\begin{figure*}
    \begin{tabular*}{\textwidth}{M{0.13\textwidth}M{0.39\textwidth}M{0.39\textwidth}}
        \hline
        \diagbox[width=8em]{Parameter}{Dataset} & MNIST & FMNIST \\
        \hline
        Number of patches &  \includegraphics[width=55mm]{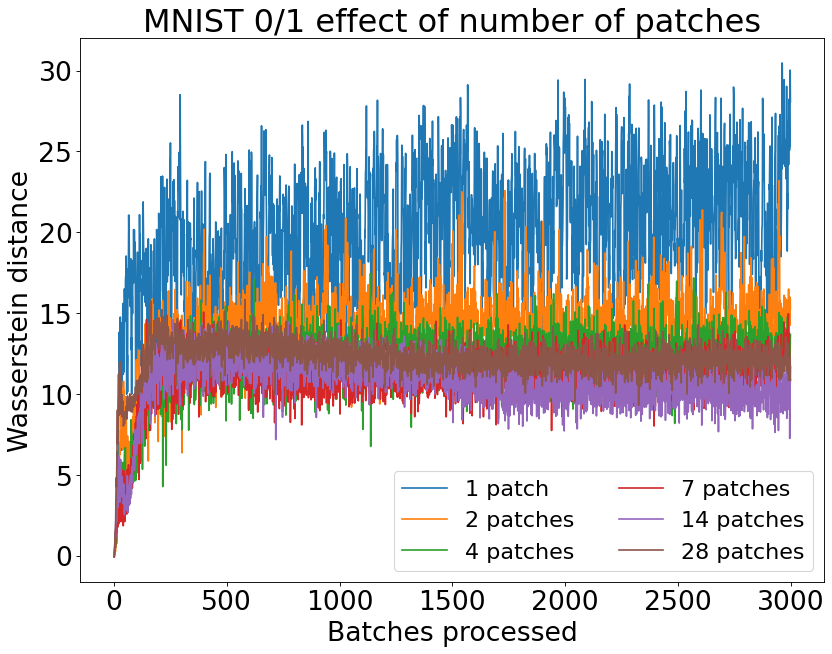} & \includegraphics[width=55mm]{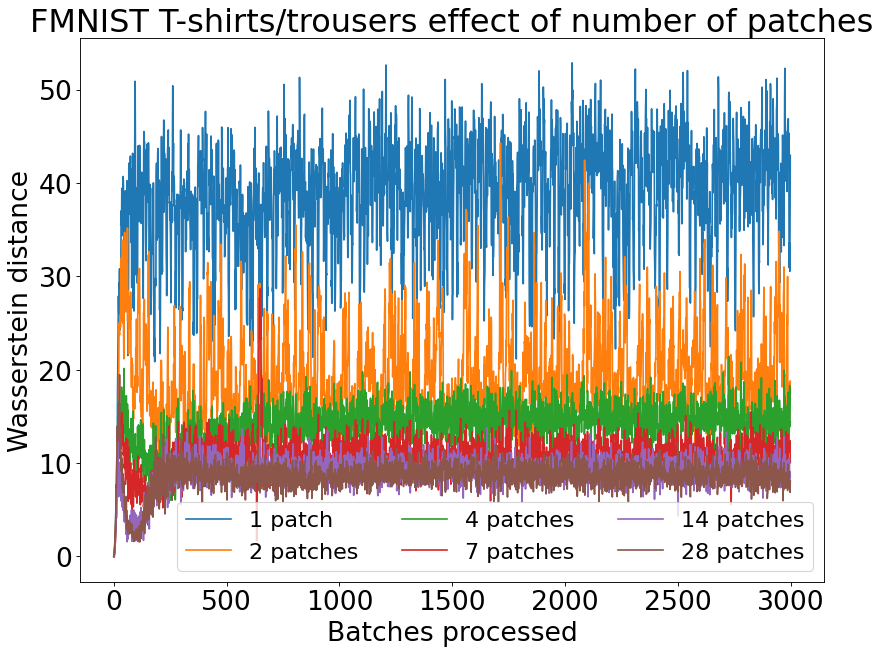} \\
        Number of qubits & \includegraphics[width=55mm]{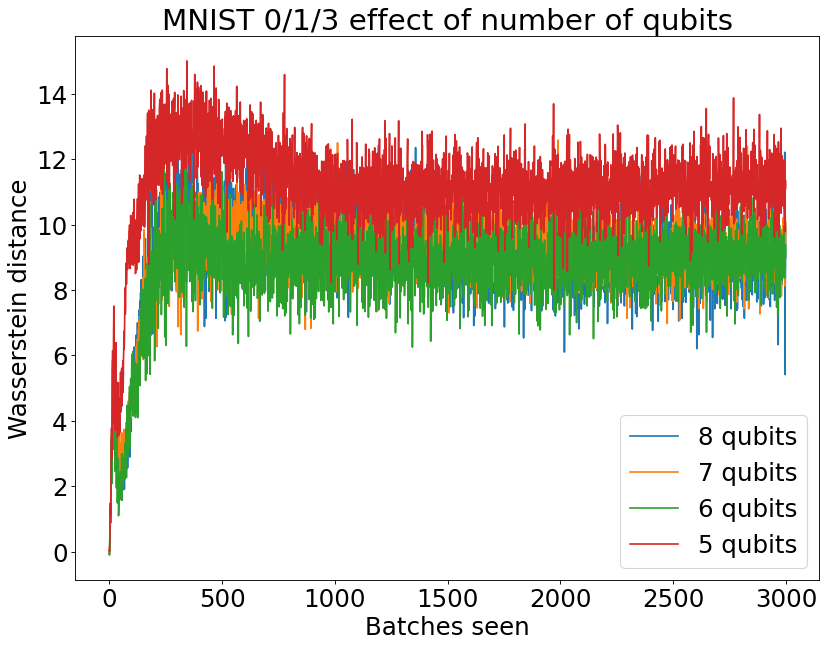} & \includegraphics[width=55mm]{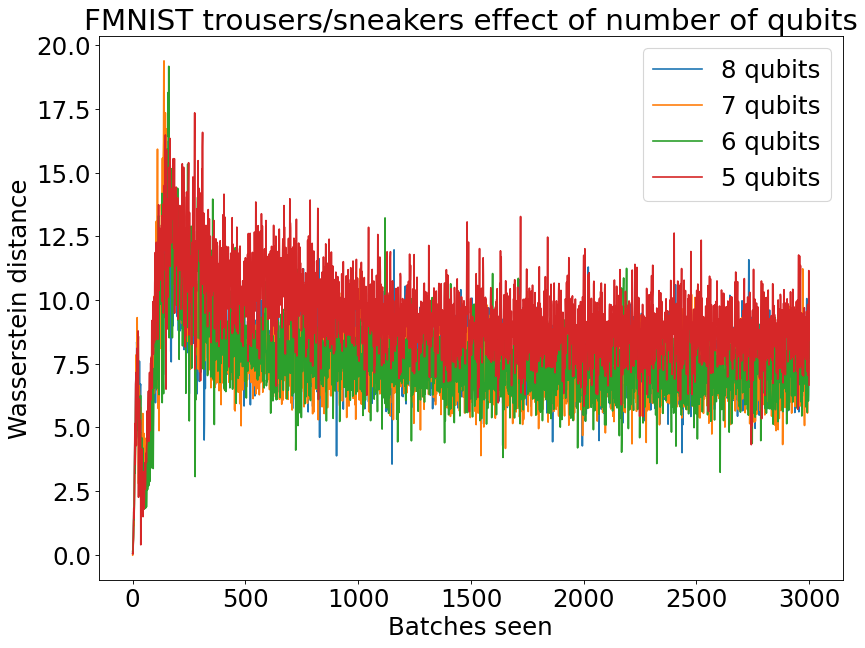} \\
        Number of layers & \includegraphics[width=55mm]{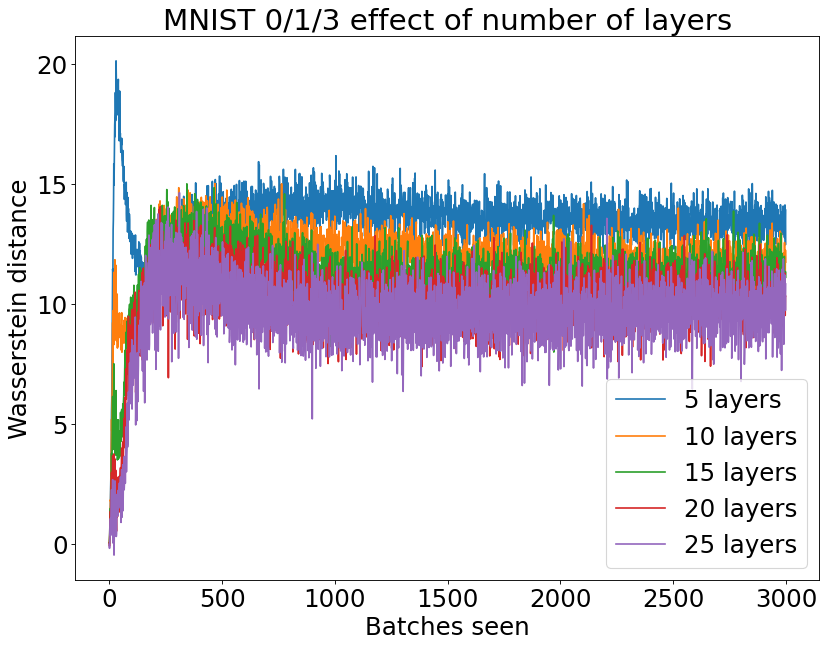} & \includegraphics[width=55mm]{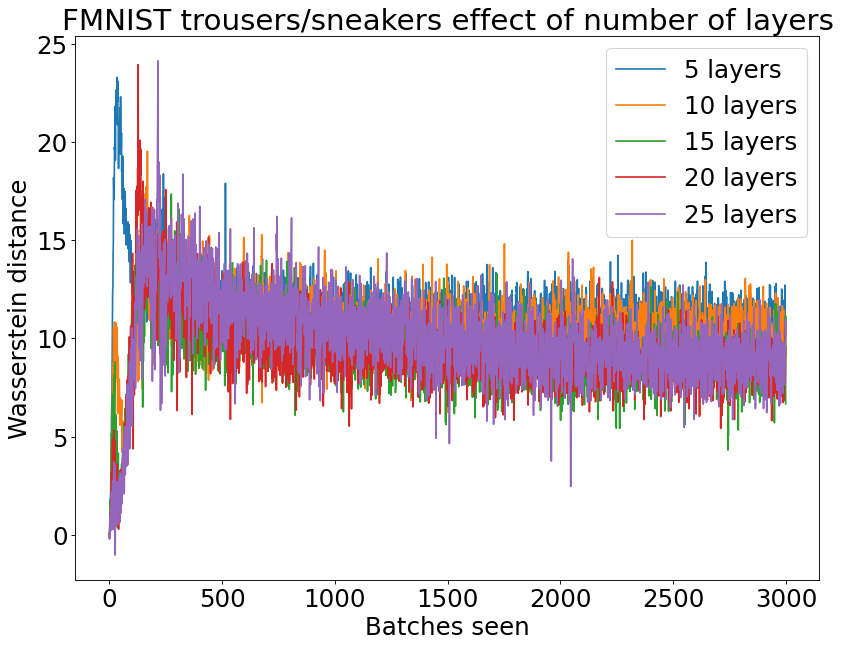} \\
        Shape of patches & \includegraphics[width=55mm]{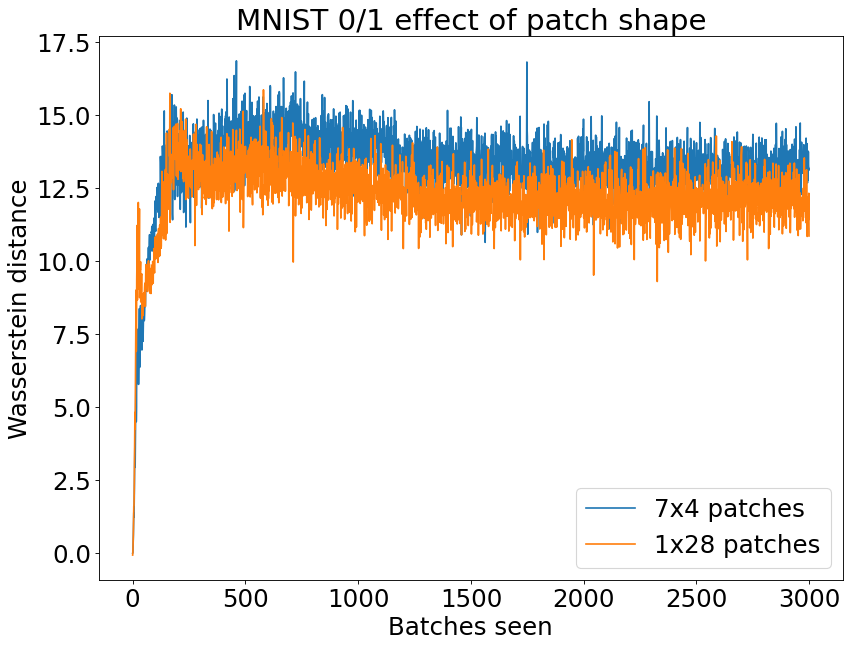} & \includegraphics[width=55mm]{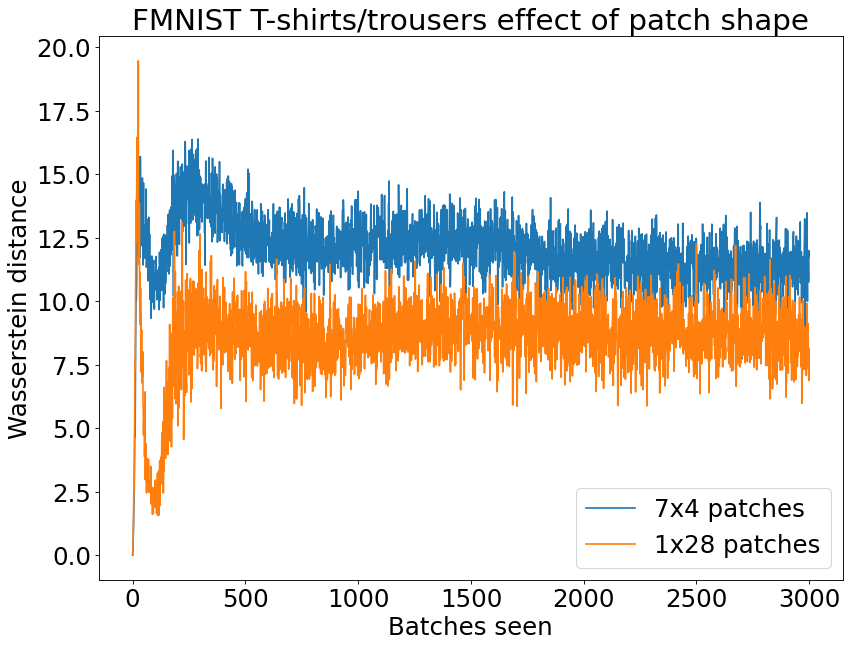} \\
        Prior distribution & \includegraphics[width=55mm]{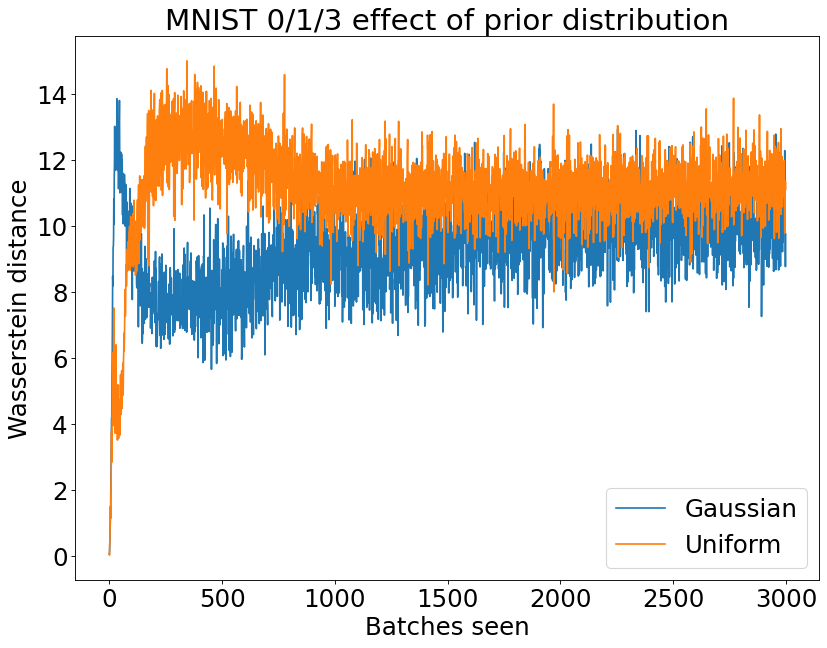} & \includegraphics[width=55mm]{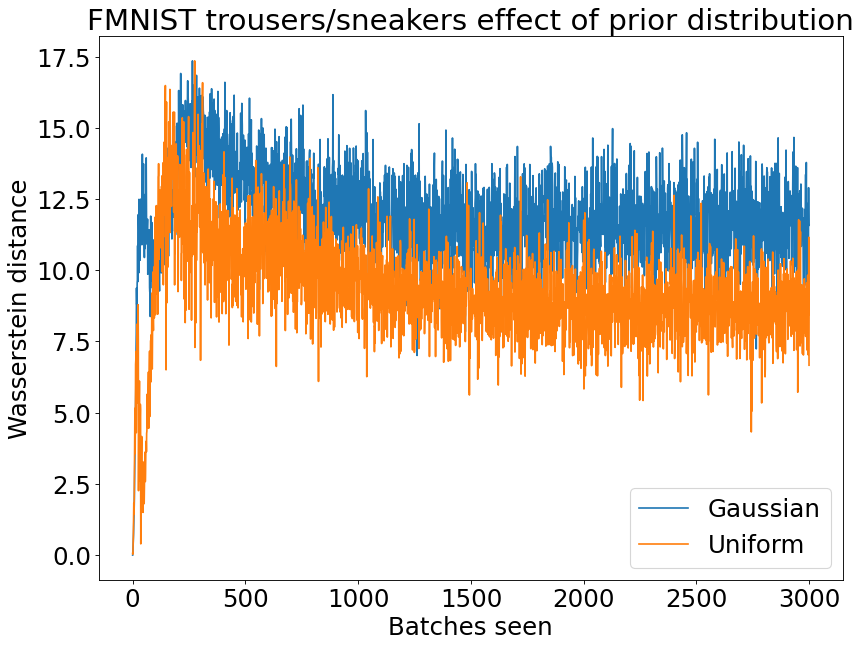} \\
        \hline
    \end{tabular*}
    \caption{Collection of the training curves from the parameter experiments conducted in Section \ref{section:effects}.}
    \label{fig:training-curves}
\end{figure*}

A collection of the training curves from the parameter experiments conducted in Section \ref{section:effects} can be found in Fig. \ref{fig:training-curves}.

\section*{Acknowledgements}

The authors acknowledge useful feedback from Amena Khatun (CSIRO). MTW acknowledges the support of the Australian Government Research Training Program Scholarship. SME is in part supported by Australian Research Council (ARC) Discovery Early Career Researcher Award (DECRA) DE220100680. The work was partially supported by the Australian Army through Quantum Technology Challenge 2022. The computational resources were provided by the National Computing Infrastructure (NCI) and Pawsey Supercomputing Center through National Computational Merit Allocation Scheme (NCMAS), and also by The University of Melbourne’s Research Computing Services and the Petascale Campus Initiative.
\\ \\
\noindent
\textbf{Data availability:}
The data that support the findings of this study are available within the article.
\\ \\
\noindent
\textbf{Code availability:}
The code to run the simulations can be found at https://github.com/jasontslxd/PQWGAN.
\\ \\
\noindent
\textbf{Competing financial interests:} 
The authors declare no competing financial or non-financial interests.

\bibliographystyle{naturemag}

\end{document}